\documentclass{article}

\usepackage{amsthm}
\usepackage{amsmath}
\usepackage{graphicx}
\usepackage{amsmath}
\usepackage{graphicx}
\usepackage{amssymb}
\usepackage{here}
\usepackage{lscape}
\usepackage{amsfonts}
\usepackage{mathtools}
\usepackage{enumerate}
\usepackage[hidelinks]{hyperref}
\usepackage{authblk}
\usepackage[british]{babel}
\usepackage{pifont}

\usepackage{soul}
\usepackage{xcolor}

\usepackage{rotating}
\usepackage{booktabs}
\usepackage[flushleft]{threeparttable}
\usepackage{blindtext}
\usepackage{geometry}
\newtheorem{myprop}{Proposition}
\newtheorem{mytheo}{Theorem}
\newtheorem{mylemma}{Lemma}
\newtheorem{mycorollary}{Corollary}

\let\oldint\int
\renewcommand{\int}{\oldint\limits}

\let\oldsum\sum
\renewcommand{\sum}{\oldsum\limits}
\begin{document}
\title{Maximum likelihood estimation for left-truncated log-logistic distributions with a given truncation point}
\author[1]{Markus Kreer}
\affil[1]{Feldbergschule, Oberh\"ochstadter Str. 20, 61440 Oberursel (Taunus), Germany.}

\author[2]{Ay{\c s}e K{\i}z{\i}lers\"u}
\affil[2]{CSSM, Department of Physics, University of Adelaide, 5005,
Adelaide, Australia.} 

\author[2]{Jake Guscott}

\author[3]{Lukas Christopher Schmitz}
\affil[3]{Institut f\"ur Mathematik, Johannes Gutenberg-Universit\"at Mainz, Staudingerweg 9, 55128 Mainz, Germany.}

\author[2]{Anthony W. Thomas}


\maketitle

\begin{abstract}
The maximum likelihood estimation of the left-truncated log-logistic distribution with a given truncation point is analyzed in detail from both mathematical and numerical perspectives. These maximum likelihood equations often do not possess a solution, even for small truncations. A simple criterion is provided for the existence of a regular maximum likelihood solution. In this case a profile likelihood function can be constructed and the optimisation problem is reduced to one dimension.
When the maximum likelihood equations do not admit a solution for certain data samples, it is shown that the Pareto distribution is the $L^1$-limit of the degenerated left-truncated log-logistic distribution. 
Using this mathematical information, a highly efficient Monte Carlo simulation is performed to obtain critical values for some goodness-of-fit tests. The confidence tables and an interpolation formula are provided and several applications to real world data are presented. 
\end{abstract}

\section{Preliminaries}

The log-logistic distribution, also known as the Fisk distribution, has been popular in the econometric community since the early 1960s because of its better description of income distributions (\cite{Fisk1961}), as compared with the Pareto distribution. On the other hand, hydrologists in the late 1980s suggested that the log-logistic distributions were useful for modelling Canadian precipitation data (\cite{Shoukri1988}), or flood frequencies in Scotland for annual flood maxima (\cite{Sinclair1988}). The log-logistic distribution is related to the logistic distribution by a logarithmic transform. Practioners make use of log-logistic distributions  because of its easy calculability and closed form expressions for both cummulative probability distribution (cdf) and probablity density function (pdf) \cite{Reath2018},\cite{He2020}. For $x>0$ the pdf and cdf are given by
\begin{eqnarray}
f(x|\alpha,\beta) & = & \frac{\beta}{\alpha} \left(\frac{x}{\alpha}\right)^{\beta-1} \frac{1}{\left[1+(x/\alpha)^{\beta}\right]^2}
\nonumber
\\
F(x|\alpha,\beta) & = & \frac{1}{1+(x/\alpha)^{-\beta}}
\nonumber
\end{eqnarray}
where $\alpha>0$ is the scale parameter and $\beta>0$ the shape parameter. We can recover the Pareto distribution for the tail by expanding the cdf for ``large'' arguments, $x \gg \alpha$, because up to first order $F(x) \simeq 1-(x/\alpha)^{-\beta}$.

In the analysis of annual flood maxima, the  log-logistic distribution in Ref.~\cite{Sinclair1988} was modified by the introduction of a threshold parameter for practical reasons: a flood maximum in a rainy country like Scotland should always be above a certain threshold level. In this paper we pursue a different approach: we keep the two-parameter distribution but introduce a fixed left-truncation point $x_L>0$ instead. Thus we have now for $x>x_L$ the left-truncated log-logistic pdf and cdf respectively as (see also Refs.~\cite{Kendall1979,Cohen1991,Guscott2018} for obtaining a truncated distribution from a complete distribution)
\begin{eqnarray}
f_{LT}(x|\alpha,\beta;x_L) & = & \left(1+ \left( \frac{x_L}{\alpha} \right)^{\beta}\right) \frac{\beta}{\alpha} \left( \frac{x}{\alpha} \right)^{\beta-1}  \frac{1}{\left[1+\left(\frac{x}{\alpha}\right)^{\beta}\right]^2} \quad,
\label{eq:pdf_trunc}
\\
 F_{LT}(x|\alpha,\beta;x_L) &=& 
\frac{\left(\frac{x}{\alpha}\right)^{\beta}- \left( \frac{x_L}{\alpha} \right)^{\beta}}{1+\left(\frac{x}{\alpha}\right)^{\beta}} \quad.
\label{eq:cdf_trunc}
\end{eqnarray}
The subscript ``LT" stands for left-truncated.If a random variable $X$ is log-logistically distributed with positive parameters $\alpha, \beta$ and left-truncation $x_L$ we can denote $X \sim LL(\alpha,\beta;x_L)$. From Eq.~(\ref{eq:cdf_trunc}) we see immediately how to generate a random variable $X \sim LL(\alpha,\beta;x_L)$ from a uniformly distributed random variable $U$ in the interval $(0,1)$, namely
\begin{eqnarray}
X = \left( \frac{\alpha^{\beta} U + x_L^{\beta}}{1-U} \right)^{1/\beta} = \alpha \left( \frac{ U + \eta}{1-U} \right)^{1/\beta} \quad,
\label{eq:Xll_trunc}
\end{eqnarray}
where the $\eta = (x_L/\alpha)^{\beta}$.

When probability distributions are truncated some interesting effects can happen: in \cite{Castillo1994} it is demonstrated that if a normal distribution is truncated, there exists finite random samples for which the regular maximum likelihood equations (MLE) do not possess a solution. Instead a new maximum likelihood estimator as a limit case was obtained leading to a degenerated one-parameter distribution, namely the exponential distribution``to fit'' the sample data appropriately. Similar effects were observed in the analysis of the left-truncated Weibull distributions \cite{KreerKizilersuThomasReis2015}, where for certain samples the MLE do not possess a solution. Indeed, in \cite{Guscott2018} it is found numerically that for a rather large set of data samples the MLE for the left-truncated log-logistic distribution do not possess a solution. 

This was one of the main motivations for this paper, because it demonstrates a clear need for a careful analysis of the left-truncated log-logistic distribution from a rigorous mathematical point of view. To the best of our knowledge there was no study to prove the existence of the maximum likelihood estimator for a random sample drawn from the left-truncated log-logistic distribution. Therefore any numerical studies without this proof, assuming the existence of a solution, will end up in the worst case either not converging at all, or converging to a degenerate solution in which the parameter estimates can take values like zero or infinity.

Our paper is structured as follows: Section 2 contains the main theorems for the existence of a non-trivial solution of the maximum likelihood equations for the left-truncated log-logistic distribution. It also examines the properties of a suitable profile likelihood function. These results  will be relevant for the efficient numerical implementation explained in section 3. Moreover we also discuss in this section our findings for the critical values for Kolmogorov-Smirnov (KS) and Anderson-Darling (AD) hypothesis tests based on extensive Monte Carlos simulations at the supercomputer of Adelaide University. As an illustration, in Section 4 we apply our technique to cancer data  and German precipitation data. All proofs are given in Section 5.

\section{Mathematical results}

\subsection{Scaling property}
We first provide a lemma dealing with a scaling property.

\begin{mylemma}(Scaling Property)
Let $X \sim LL(\alpha,\beta;x_L)$ be a left-truncated log-logistic random variable. Then for any $k>0$ we have $kX \sim LL(k\alpha,\beta;k x_L)$.
\end{mylemma}

 {\em Proof:}

By assumption $X>x_L$ with $X \sim LL(\alpha,\beta;x_L)$. Thus $kX>k x_L$ and

\begin{eqnarray}
\text{prob}(kX<x~|~kX>k x_L) & = & \text{prob}(X<x/k| X> x_L)
\nonumber \\
& = & \frac{\left(\frac{x/k}{\alpha}\right)^{\beta}- \left( \frac{x_L}{\alpha} \right)^{\beta}}{1+\left(\frac{x/k}{\alpha}\right)^{\beta}} 
 =  \frac{\left(\frac{x}{k\alpha}\right)^{\beta}- \left( \frac{k x_L}{k\alpha} \right)^{\beta}}{1+\left(\frac{x}{k\alpha}\right)^{\beta}}
\nonumber
\end{eqnarray}
and the proof is finished. $\blacksquare$

As a consequence, for simplicity we will assume later on the left-truncation point $x_L=1$. In other words we rescale the independent identically distributed (i.i.d.) sample $X_1,...,X_N$ with $k=1/x_L$, leading to $X_1/x_L,...,X_N/x_L$  (which is truncated at $1$). In any stage we can go back to the original sample and original parameters by using a rescaling factor $k=x_L$.

\subsection{The first-order maximum likelihood equations}

We introduce a new parametrization with $\lambda=\alpha^{\beta}$, and re-write Eq.~(\ref{eq:pdf_trunc}) as
\begin{eqnarray}
f_{LT}(x|\lambda,\beta) & = & \left(1+ \frac{x_L^{\beta}}{\lambda}\right) \frac{\beta}{\lambda} x^{\beta-1} \frac{1}{\left[1+\frac{x^{\beta}}{\lambda}\right]^2}\quad.
\label{eq:pdf_trunc_lambda}
\end{eqnarray}
Note that with this notation we have $\eta = x_L^{\beta}/\lambda$.

Using Eq.~(\ref{eq:pdf_trunc_lambda}) and a left-truncated log-logistic sample, denoted by $X_1, X_2, ..., X_N$ and all observations are bigger than $x_L>0$,  the log-likelihood function is given via by
\begin{eqnarray}
\ln{L_{LT}(\{X_i\}|\lambda,\beta;x_L)} 
& = & N \ln{ \left(1+\frac{x_L^{\beta}}{\lambda}  \right)} + N \ln{\beta} - N \ln{\lambda}
\nonumber \\
& & + (\beta-1) \sum_{i=1}^{N}  \ln{X_i} 
-2 \sum_{i=1}^{N}  \ln{ \left[ 1+ \frac{X_i^{\beta}}{\lambda}   \right] } \, .
\label{eq:ML2}
\end{eqnarray}
The MLE equations are obtained by differentiating Eq.~(\ref{eq:ML2}) with respect to $\lambda$ 
and $\beta$ respectively and putting the derivatives equals to zero. From $\frac{\partial}{\partial \lambda} \ln{L_{LT}}=0$ we obtain for $(\lambda,\beta)\in  \mathbb{R^+}\times \mathbb{R^+}$
\begin{eqnarray}
0 & = & \frac{2}{N}  \sum_{i=1}^{N}  \frac{ \frac{X_i^{\beta}}{\lambda}}{1+\frac{X_i^{\beta}}{\lambda}}   -
1-\frac{ \frac{x_L^{\beta}}{\lambda}}{1+\frac{x_L^{\beta}}{\lambda}}
\label{eq:lambda}
\end{eqnarray}
and from $\frac{\partial}{\partial \beta} \ln{L_{LT}}=0$ we obtain for $(\lambda,\beta)\in \mathbb{R^+}\times \mathbb{R^+}$
\begin{eqnarray}
0 & = &  \frac{  \frac{x_L^{\beta}}{\lambda} \ln{ x_L}}{1+\frac{x_L^{\beta}}{\lambda}} +    
\frac{1}{\beta} + \frac{1}{N}\sum_{i=1}^N \ln{X_i} - \frac{2}{N}  \sum_{i=1}^{N}  \frac{ \frac{X_i^{\beta}}{\lambda}}{1+\frac{X_i^{\beta}}{\lambda}}\ln{X_i} \, .
\label{eq:beta_new}
\end{eqnarray}
A solution  of these  MLE equations Eqs.~(\ref{eq:lambda})--(\ref{eq:beta_new}) (if it exists) will be denoted by $(\hat{\lambda},\hat{\beta})$.
For finite untruncated samples, Ref.~\cite{Gupta1999} (respectively Ref.~\cite{Antle1970}) has shown that the MLE have a unique solution for the log-logistic distribution (respectively the logistic distribution). These proofs fail when a left-truncation point $x_L>0$ is introduced.

We will take $x_L=1$ from now on without loss of generality. We also define for convenience the quantity $S=\sum_{i=1}^N \ln{X_i}$ and introduce a new objective function as in~\cite{Gupta1999}
\begin{eqnarray}
\varphi(\lambda,\beta) & = & \ln{L_{LT}(\{X_i\}|\lambda,\beta;x_L=1)} + S \quad.
\nonumber
\end{eqnarray}
The extrema of both the function $\varphi(\cdot,\cdot)$ and the log-likelihood function are the same (and so are their MLE equations), because they differ by a constant number, $S$.

\subsection{Our main theorems}
Our first theorem states the existence of a maximum for the objective function under certain conditions.
\begin{mytheo}{\label{thm:exist}}(Existence)
Consider the i.i.d. left-truncated sample $\protect{X_1,..., X_N >1}$, for which at least two observations are different, and define the objective function 
$\varphi(\cdot,\cdot): \mathbb{R^+}\times \mathbb{R^+}\rightarrow  \mathbb{R}$ by
\begin{eqnarray}
\varphi(\lambda,\beta) 
& = &  N \ln{ \left(1+\frac{1}{\lambda}  \right)} + N \ln{\beta} - N \ln{\lambda} + \beta S
\nonumber \\
& & 
-2 \sum_{i=1}^{N}  \ln{ \left[ 1+ \frac{X_i^{\beta}}{\lambda}   \right] } \, .
\label{eq:objective}
\end{eqnarray}
Define $\beta_C>0$ as the unique solution of 
\begin{eqnarray}
\frac{1}{2} = \frac{1}{N} \sum_{i=1}^N \frac{1}{X_i^{\beta_C}}
\label{eq:betastar}
\end{eqnarray}
and $\beta_0>0$ by
\begin{eqnarray}
\frac{1}{\beta_0} =  \frac{1}{N}\sum_{i=1}^{N} \ln{X_i}\quad.
\label{eq:beta0}
\end{eqnarray}

Then the following holds true
\begin{enumerate}[{(1)}]
\item For  $\beta_0>\beta_C$ the objective function $\varphi(\cdot,\cdot)$ posseses a global maximum $(\hat{\lambda},\hat{\beta}) \in \mathbb{R^+}\times \mathbb{R^+}$.

\item For $\beta_0 \leq \beta_C$ the objective function $\varphi(\cdot,\cdot)$ posseses a (local) maximum $(\hat{\lambda},\hat{\beta}) =(0,\beta_0)$ on the boundary.
\end{enumerate}
\end{mytheo}
 The proof is given in Section \ref{sec:proofs}.
 
The next result is important for the numerical computation of the maxima and leads to a profile likelihood function and provides a curve for the loci\footnote{ locus = a set of points that satisfy or are determined by some specific condition} of critical points.

\begin{mytheo}{\label{thm:profile}}(Loci of critical points)

Consider the i.i.d. left-truncated sample $X_1,..., X_N >1$ for which at least two observations are different. Then the following holds true:
\begin{enumerate}[{(1)}]
\item For fixed $\beta >\beta_C$ the equation
\begin{eqnarray}
\left. \frac{\partial}{\partial \lambda}\varphi(\lambda,\beta) \right|_{\lambda=\hat{\lambda}} = 0
\nonumber
\end{eqnarray}
has exactly one positive solution $\hat{\lambda}$ which depends on the fixed parameter $\beta$. Furthermore we have the following inequalities
\begin{eqnarray}
\frac{\partial}{\partial \lambda}\varphi(\lambda,\beta) &>& 0 \hspace{0.5 cm}\text{for} \hspace{0.5 cm} 0<\lambda <\hat{\lambda}
\nonumber \\
\frac{\partial}{\partial \lambda}\varphi(\lambda,\beta) &<& 0 \hspace{0.5 cm}\text{for} \hspace{0.5 cm} \lambda >\hat{\lambda}
\nonumber
\end{eqnarray}
and thus
\begin{eqnarray}
\left. \frac{\partial^2}{\partial \lambda^2}\varphi(\lambda,\beta) \right|_{\lambda=\hat{\lambda}} < 0\quad.
\label{eq:min_lambda}
\end{eqnarray}
 \item For fixed $0<\beta \leq \beta_C$ the equation
\begin{eqnarray}
\left. \frac{\partial}{\partial \lambda}\varphi(\lambda,\beta) \right|_{\lambda=\hat{\lambda}} = 0 \quad,
\nonumber
\end{eqnarray}
has the only solution $\hat{\lambda}=0$. 
\end{enumerate}
Hence the non-negative continous function $\Lambda(\cdot): \mathbb{R}^{+}\rightarrow \mathbb{R}^{+}_{0}$, defined by
\begin{eqnarray}
\Lambda(\beta)
= \left\{\begin{array}{ll} 0, & 0<\beta \leq \beta_C \\
\hat{\lambda}, & \beta >\beta_C \end{array}\right.
\label{eq:Lambda_beta}
\end{eqnarray}
is the locus of all possible critical points of the objective function $\varphi(\cdot,\cdot)$.
\end{mytheo}
The proof is given in Section \ref{sec:proofs}.

By inserting the function $\Lambda(\beta)$, as constructed in Eq.~(\ref{eq:Lambda_beta}), into the original objective function Eq.~(\ref{eq:objective}) we obtain the ``profile likelihood function'' 
\begin{eqnarray}
\tilde{\varphi}(\beta)=\varphi(\Lambda(\beta),\beta)
&=& \left\{\begin{array}{ll} N\ln{\beta}-\beta S, & \beta \in (0,\beta_C] \\ \\
        N\ln{\left(1+\frac{1}{\Lambda(\beta)}\right)}+ N\ln{\beta} -N\ln{\Lambda(\beta)}    & \nonumber\\
+ \beta S- 2 \sum_{i=1}^N \ln{\left(1+\frac{X_i^{\beta}}{\Lambda(\beta)}\right)}, & \beta \in (\beta_C,\infty)\end{array}\right.
\nonumber \\
& &
\label{eq:profilelikelihood}
\end{eqnarray}

\begin{figure}[hptb]
	\begin{center}
		\includegraphics[width=0.7\textwidth]{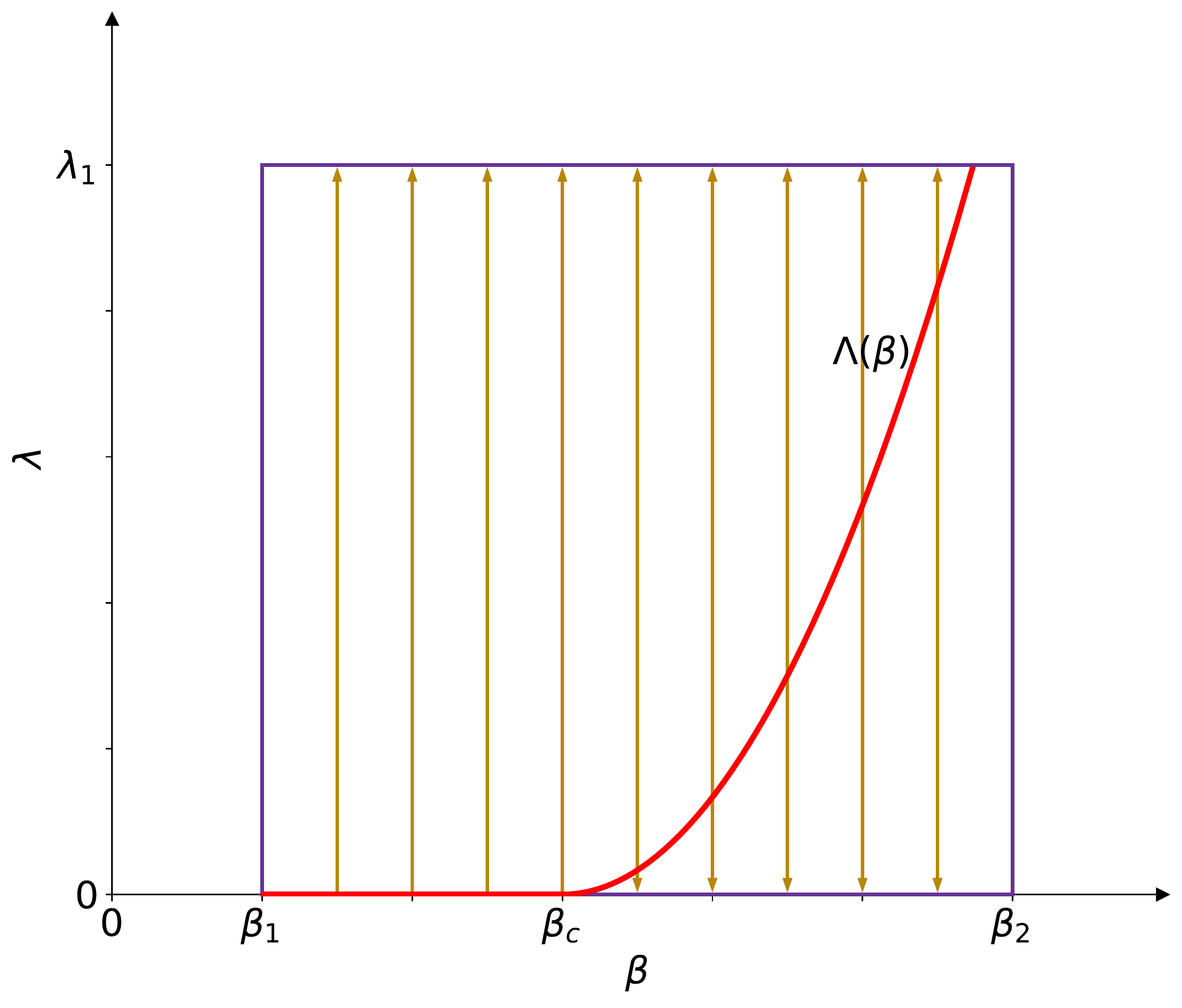}
		\caption{ The function $\Lambda(\beta)$ for the profile likelihood.\label{fig:function_Lambda_beta}}
	\end{center}
\end{figure}

Therefore, we have reduced the two-dimensional maximization problem for $\varphi(\cdot,\cdot)$ on a finite region to a one-dimensional problem for $\tilde{\varphi}(\cdot)$ on a finite interval. 
Figure~\ref{fig:function_Lambda_beta} illustrates the problem: the critical points (and therefore our maximum guaranteed by Theorem~\ref{thm:exist})  are located inside the blue rectangle region and must be located on the red curve. By condition Eq.~(\ref{eq:min_lambda}), critical points  can only be maxima or saddle points. This simplification is important because it improves the speed of finding the critical points in our numerical study dramatically. Unfortunately, we were not able to prove uniqueness for the critical points, as the necessary concavity arguments for the Hessian matrix seem intractable. Numerically, though, we always found exactly one critical point for millions of samples generated and this point was always a maximum.

A Corollary sumarises the situation of our two theorems.

\begin{mycorollary}{\label{thm:master}}
Under the assumptions of Theorem~\ref{thm:exist}. and \ref{thm:profile}. the following holds true
\begin{enumerate}[{(1)}]
\item If and only if $\beta_0>\beta_C$ the MLE equations Eq.~(\ref{eq:lambda})--(\ref{eq:beta_new}) possess a solution $(\hat{\lambda},\hat{\beta})\neq (0,0)$ (which might not be unique). 
Critical points are obtained by constructing the function $\Lambda(\cdot):  \mathbb{R}^{+}\rightarrow  \mathbb{R}^{+}_{0}$ as defined for all $\beta>0$ in Theorem~\ref{thm:profile},  Eq.~(\ref{eq:Lambda_beta}), and then solving Eq.~(\ref{eq:beta_new}), which now reads
\begin{eqnarray}
0 = \frac{N}{\beta}+\sum_{i=1}^N \ln{X_i}-2\sum_{i=1}^N\frac{X_i^{\beta}\ln{X_i}}{\Lambda(\beta)+X_i^{\beta}}\quad,
\label{eq:master}
\end{eqnarray}
for the desired critical point $\hat{\beta}>\beta_C$.
\item If and only if $\beta_0\leq\beta_C$ the MLE equations, Eq.~(\ref{eq:lambda})--(\ref{eq:beta_new}), do not possess a solution. The likelihood attains a maximum on the boundary 
$(\hat{\lambda},\hat{\beta})= (0,\beta_0)$ with a Pareto probabilty distribution given by
\begin{eqnarray}
g(x|\beta_0) = \frac{\beta_0}{x^{1+\beta_0}}\quad.
\label{eq:Pareto_beta0}
\end{eqnarray}
\end{enumerate}
\end{mycorollary}
The proof is given in Section \ref{sec:proofs}.

Our final theorem demonstrates that the condition that at least two observations in our sample be different is not only sufficient but also necessary. We see that in this case no finite maximum of the MLE  exists.

\begin{mytheo}{\label{thm:identical}}(Necessity of at least two different observations)
Consider the i.i.d. left-truncated sample $X_1=X_2=...=X_N >1$ for which all observations are equal to $X_1$, say. Then the following holds true:
\begin{enumerate}[{(1)}]
\item $\beta_0=1/\ln{X_1}$ and $\beta_C=\ln{2}/\ln{X_1}$

\item For fixed $\beta >\beta_C$ the function
\begin{eqnarray}
\Lambda(\beta)=X_1^{\beta}-2
\nonumber
\end{eqnarray}
satisfies the ML equation for $\lambda$, Eq.~(\ref{eq:lambda}).

\item The profile likelihood function Eq.~(\ref{eq:profilelikelihood}) reading here as
\begin{eqnarray}
\tilde{\varphi}(\beta)=\varphi(\Lambda(\beta),\beta)
=\left\{\begin{array}{ll} N\ln{\beta}- N\beta \ln{X_1}, & \beta \in (0,\beta_C] \\ \\
        -N\ln{\left(X_1^{\beta}-1\right)}+ N\ln{\beta}     & \nonumber\\
+ N\beta \ln{X_1} - 2 N \ln{2}, & \beta \in (\beta_C,\infty)\end{array}\right.
\end{eqnarray}
is continous and strictly monotonically increasing in $\beta$. It therefore does not possess a finite maximum.
\end{enumerate}
\end{mytheo}
The proof is given in Section \ref{sec:proofs}.

This last Theorem, which can be also formulated in the same fashion for the un-truncated log-logistic MLE,  might explain the phenomenon observed in Ref.~\cite{Shoukri1988}, Table 1, p. 231: there the percentage of failures of the maximisation algorithm for the likelihood is given for small samples. Indeed, when a sample consists of e.g. $N=15$ data points which are all ``nearly equal'', a direct optimiziation by numerical methods is bound to fail if the parameter region chosen is too ``small''. The profile likelihood will be increasing in one direction towards the boundaries of the ``small'' parameter region and the Newton-Raphson solver of the MLE equations fails to converge. That the MLE problem in this case possesses a solution nevertheless was established in Ref.~\cite{Gupta1999}.  


\section{Monte Carlo study and statistical hypothesis test}

The mathematical results obtained in the previous section are now applied in the framework of hypothesis testing. The determination of critical values for a certain confidence level are necessary in the case that the distribution parameters are estimated from the given sample, ``in-sample testing". Some of us have shown in the case of left-truncated Weibull distribution that the critical values will depend on sample size, the truncation point and also on the parameters. This functional dependence will be investigated now in detail. Extensive Monte Carlos studies have been performed in \cite{Guscott2018} to determine the functional dependence. In this paper we present the results for the Kolmogorov-Smirnov and Anderson-Darling hypothesis tests for various confidence levels.

\subsection{Kolmogorov-Smirnov and Anderson-Darling hypothesis test}{\label{tests}}
If the MLE for a given sample of size $N$ yields non-trivial parameters $(\hat{\lambda},\hat{\beta})$, then
the Kolmogorov-Smirnov (KS) distance $D$ for the sample s, defined as being the maximum value of the absolute difference between the empirical distribution function and the cummulative distribution function $F(\cdot|\hat{\lambda},\hat{\beta})$ (see ~\cite{Kolmogorov1933,Smirnov1948}) is
\begin{eqnarray}
D = \max_{1\leq i \leq N} { \left\{ \frac{i}{N}-F(X_i|\hat{\lambda},\hat{\beta}),\frac{i-1}{N}-F(X_i|\hat{\lambda},\hat{\beta}) \right\}} \, .
\label{eq:KS}
\end{eqnarray}
We employ here as test statistic for the KS test the quantity $\sqrt{N}D$.

The Anderson-Darling (AD) test statistic $A^2$ is given by the following expression (\cite{AndersonDarling1952})
\begin{eqnarray}
A^2= -N-\frac{1}{N}\sum_{i=1}^N \{2i-1\}\left\{\ln{[F(X_i|\hat{\lambda},\hat{\beta})]}+\ln{[1-F(X_{N-i+1}|\hat{\lambda},\hat{\beta})]} \right\}\quad.
\label{eq:AD}
\end{eqnarray}

If these positive test statistics are smaller than certain critical values (depending on the confidence level) we cannot reject the statement that ``the sample data are distributed according to a truncated log-logistic distribution with estimated parameters $(\hat{\lambda},\hat{\beta})$'' for a given confidence level. 

\subsection{Algorithmic implementation and results}
The following Monte Carlo algorithm to determine critical numbers for the KS and AD tests was performed in ~\cite{Guscott2018}. The maximum likelihood estimation in the spirit of 
Corollary~\ref{thm:master} is algorithmically very efficient and despite the fact that we could not assert uniqueness, it turns out that numerical search for critical points seems to yield always exactly one solution which satisfies the properties of a maximum. The flow chart of the algorithm for a given sample size is as follows
\begin{enumerate}
\item For a fixed integer $N$ iterate the following steps 2 to 5, $C$ times ($C$ being an integer). 

\item Draw random sample of log-logistic numbers of size $N$ for given parameters $\lambda, \beta$ for fixed truncation $x_L=1$ using Eq. (\ref{eq:Xll_trunc}) and $\lambda=\alpha^{\beta}$.

\item Calculate  $\beta_C$ from Eq.~(\ref{eq:betastar}) and  $\beta_0$ from Eq.~(\ref{eq:beta0}). If $\beta_0\leq \beta_C$ then the critical point corresponding to a solution to the maximum likelihood estimation problem is a Pareto distribution with $\hat{\lambda}=0$ and $\hat{\beta}=\beta_0$ and finish at this step, otherwise go to step 3.

 \item If $\beta_0>\beta_C$, then solve the equation Eq.~(\ref{eq:master}) to obtain $\hat{\beta}$ 
and $\hat{\lambda}=\Lambda(\hat{\beta})$ via Eq. (\ref{eq:Lambda_beta})

\item If $\hat{\lambda}>0$, compute KS and AD statistic like in Section \ref{tests}, Eq.~(\ref{eq:KS}) and Eq.~(\ref{eq:AD}) respecitvely and store values in lists and go back to 1. 

\item Sort list of KS and AD values in ascending order and find quantile in usual way \cite {Guscott2018}.
\end{enumerate}
This has been repeated for various sample sizes.

The subsequent results have been obtained  in ~\cite{Guscott2018}. Following the notation therin and using our parameter $\eta$ from above, which is either zero in the case of $x_L=0$ (untruncated case) or otherwise $\eta=1/\lambda$ with the normalization of $x_L=1$, the quantity $p = F(x_L~|~\alpha,\beta)$ is introduced as a measure for the ``truncation percentage'', reading here as
\begin{eqnarray}
p = 1-\frac{1}{1+\eta}=1-\frac{1}{1+\lambda^{-1}} \, .
\nonumber
\end{eqnarray}
This truncation percentage tells us how many data have been ``lost'' by the left-truncation and depends, with the normalisation $x_L=1$, only on the scale parameter $\lambda$. 
Sample sizes have been chosen as $N\in\{30, 50, 100, 200, 500, 1000, 10000 \}$ and for sample size $N=10000$ altogether $C=100$ repetitions have been performed to keep $N\cdot C=1000000$.  Likewise, for $N=100$ $C=10000$ repetitions have been performed to get error estimates for the obtained quantiles (\cite{Mosteller1946,Cuddington2011}). In Tables~\ref{tab:KS95} -- \ref{tab:AD95} we display the critical values  for the 95\% confidence level for the KS and AD hypothesis tests. Additional tables for other confidence levels are provided in Appendix~\ref{sec:Tables}. Extensive power testing has been done  in Ref.~\cite{Guscott2018}. We want to emphasize here that our 95\%-KS test is able to distinguish log-logistic distributions from other fat-tailed distributions (such as Weibull, Mittag-Leffler and log-normal) only for samples with sample size $N\geq 1000$, whereas the 95\%-AD test already performs well for sample size $N\geq 100$.

We summarise our results for various confidence levels in the following interpolation formula for the test statistics $KS(\sqrt{N}D)$ and $AD(A^2)$ respectively (see \cite{Guscott2018}):
\begin{eqnarray}
\text{Test Statistics} = \frac{\theta_1 \eta + \theta_2 \eta^{1/2} +\theta_3}{\theta_4\eta^{1/2}+\theta_5+\eta}
+\theta_6 \left( \frac{\eta}{N}\right)^{1/2}+ \theta_7 \eta^{3/2} + \frac{\theta_8}{N^{1/2}}
+ \frac{\theta_9}{N}\quad.
\label{eq:interpol}
\end{eqnarray}
The values of the $\theta$'s are given in Table~\ref{tab:theta} in Appendix~\ref{sec:Tables_95_percentail}. Note that we take $\eta \simeq \hat{\eta}=1/\hat{\lambda}$ (with $x_L=1$).

Our critical values for $p=0$ may be compared with the corresponding critical values given in Ref.~\cite{Stephens1979} for the logistic distribution.

\section{Applications to real world data}

\subsection{Bladder cancer data set from Lee and Wang (2003)}

We apply our methods to the bladder cancer data set from Lee and Wang 2003~\cite{Lee2003},Table 9.3, p. 231, which has been studied recently in Ref. \cite{AlShomrani2016} under the hypothesis that the data are distributed according to a log-logistic distribution. The original Table 9.3 contains a set of 137 remission times in months from cancer patients. These remission times are subset of the data of a bladder cancer study. As in  ~\cite{AlShomrani2016}, we have used only 128 data points from this table and have taken out 9 censored observations (indicated by a ``+'' in the original Table 9.3)
\begin{eqnarray}
&& 0.08, 2.09, 3.48, 4.87, 6.94, 8.66, 13.11, 23.63, 0.20, 2.23, 3.52, 4.98, 6.97, 9.02,
\nonumber \\
&& 13.29, 0.40, 2.26, 3.57, 5.06, 7.09, 9.22, 13.80, 25.74, 0.50, 2.46, 3.64, 5.09, 7.26,
\nonumber \\
&&  9.47, 14.24, 25.82, 0.51, 2.54, 3.70, 5.17, 7.28, 9.74, 14.76, 26.31, 0.81, 2.62, 3.82, 
\nonumber \\
&& 5.32, 7.32, 10.06,14.77, 32.15, 2.64, 3.88, 5.32, 7.39, 10.34, 14.83, 34.26, 0.90, 
\nonumber \\
&& 2.69, 4.18, 5.34, 7.59, 0.66, 15.96, 36.66, 1.05, 2.69, 4.23, 5.41, 7.62, 10.75, 16.62, 
\nonumber \\
&& 43.01, 1.19, 2.75, 4.26, 5.41, 7.63,  17.12, 46.12, 1.26, 2.83, 4.33, 5.49, 7.66, 11.25,
\nonumber \\
&& 17.14, 79.05, 1.35, 2.87, 5.62, 7.87, 11.64, 17.36, 1.40, 3.02, 4.34, 5.71, 7.93, 11.79, 
\nonumber \\
&&18.10, 1.46, 4.40, 5.85, 8.26, 11.98, 19.13, 1.76, 12.07, 3.25, 4.50, 6.25, 8.37, 12.02, 
\nonumber \\
&& 2.02, 3.31, 4.51, 6.54, 8.53, 12.03, 20.28, 2.02, 3.36, 6.76, 21.73, 2.07, 3.36, 6.93,
\nonumber \\
&& 8.65, 12.63, 22.69
\nonumber
 \end{eqnarray}
Using our method we first estimate the parameters of the log-logistic distribution and then perform KS and AD hypothesis testing using the interpolation equation, Eq.~(\ref{eq:interpol}), for the critical values. This was done for the complete data set and various truncation points. Our results are displayed in Table~\ref{tab:bladder}. Note that for the purpose of comparison we use the scale parameter 
$\hat{\alpha}=\hat{\lambda}^{1/\hat{\beta}}$.

\begin{table}[htbp]
  \small
  \caption{Remission times for log-logistic distribution (95\% confidence level)}
  \begin{center}
    \begin{tabular}{| c  |  c  | c | c | c  |  c | c  |  c | c |} \hline
     $x_{L}$ [m] & $N$     &  $\hat{\alpha}$  [m]  & $\hat{\beta}$  & $\ln{L}$ & KS ($\sqrt{N}D)$    & AD $(A^2)$ &  KS & AD\\ \hline
  0    &  128     & $5.97$   & $1.695$  &  -410.89 & 0.4447  & 0.2684 & $\checkmark$ & $\checkmark$ \\ \hline
  0.25 & 126    & $6.11$   & $1.782$ &  -402.20 & 0.4344  & 0.1657 & $\checkmark$ & $\checkmark$  \\ \hline
  1    & 120     & $6.32$    & $1.877$  &  -379.28 & 0.4030  & 0.1253 & $\checkmark$ & $\checkmark$ \\  \hline
  6    & 64       & $8.63$   & $2.239$  &  -206.00 & 0.5006  & 0.3086 & $\checkmark$ & $\checkmark$ \\  \hline
  12    & 31      & $8.36$   & $2.277$   & -103.85 & 0.4877 & 0.5129 & $\checkmark$ & $\checkmark$ \\  \hline
    \end{tabular}%
  \end{center}
  \label{tab:bladder}
\end{table}%
We may compare our result without truncation with Ref.~\cite{AlShomrani2016}, which found $\hat{\alpha}=6.08982$ months and $\hat{\beta}=1.725158$, with a log-likelihood function value of $-411.4574$. These authors  were using ``LLmodel.Optim() function in R with Newton-Raphson options [...] as an iterative process for maximizing the log-likelihood function''. Our likelihood value being $-410.89$ is better and will lead to a better estimation of the parameters $\hat{\alpha}$ and 
$\hat{\beta}$. Therefore our KS distance of $(\sqrt{n}D)= 0.4447$ is a more realistic value than theirs $(\sqrt{n}D)= 0.3629$.\footnote{Using their parameter values yields exactly this KS-distance. Any difference is due to the R-optimizer.}

We used four different truncation points and estimated the parameters and performed the hypothesis testing. For all truncation points we cannot reject the hypothesis of the data being log-logistic distributed. For small truncation points  (up to ~5\% of the data) the parameter estimates remain fairly stable. Even for larger truncation points (~ 75\% of the data) our hypothesis cannot be rejected.  

\subsection{Annual precipitation for Berlin and Toronto}

In the literature the log-logistic distributions is often used to describe precipitation data, for example Refs.~\cite{Shoukri1988} and \cite{Ashkar2006}. For reason of homogeneity,~\cite{Shoukri1988} focused on precipitation data of various Canadian cities. We have chosen as an example the annual precipitation for the city-state of Berlin, because of its small area the data are likely to be i.i.d.  The data is available from the German met office~\cite{DWD2022}. The annual precipitation data consists of 141 data points starting in 1881 and extending to 2021. Also, to compare our method with Ref.~\cite{Shoukri1988}, we analyse the annual precipitation data for Toronto from 1 July to 30 June,  as given in Ref.~\cite{Toronto2022}, from 1937 until 2021.

For Berlin our results are given in Table~\ref{tab:precipitation} and for Toronto in 
Table~\ref{tab:precipitation_Toronto}. Note that in our analysis the shape parameters, $\hat{\beta}$,  for both cities are similar. We compare our results with \cite{Shoukri1988}, where for Toronto  
$\hat{\alpha}=789.8$ mm and  $\hat{\beta}=14.4$ were given. Whereas the estimates of the scale parameter $\hat{\alpha}$ are in good agreement, the values of $\hat{\beta}$ are different, but of similar order of magnitude. It is interesting to note that for Toronto a truncation is needed to pass the Anderson-Darling hypothesis test.
\begin{table}[htbp]
  \small
  \caption{Berlin precipitation data for log-logistic distribution (95\% confidence level).}
  \begin{center}
    \begin{tabular}{| c  |  c  | c | c | c  |  c | c  |  c | c |} \hline
     $x_{L}$ [mm] & $N$     &  $\hat{\alpha}$  [mm]  & $\hat{\beta}$  & $\ln{L}$ & KS $(\sqrt{N}D)$    & AD $(A^2)$ &  KS & AD\\ \hline
  0    &  141     & $564.0$   & $11.8$  &-825.30  & 0.6255  & 0.4943 & $\checkmark$ & $\checkmark$ \\ \hline
  300 & 141     & $563.9$   & $11.8$  & -825.22 & 0.6286  & 0.5001 & $\checkmark$ & $\checkmark$  \\ \hline
 400 & 137     & $565.9$   & $12.1$  & -791.86 & 0.5908  & 0.4773 & $\checkmark$ & $\checkmark$  \\ 
\hline
  500 & 113     & $574.5$   & $13.6$  & -616.60 & 0.5319  & 0.2926 & $\checkmark$ & $\checkmark$ \\  \hline
      \end{tabular}%
  \end{center}
  \label{tab:precipitation}
\end{table}
\begin{table}[htbp]
  \small
  \caption{Toronto precipitation data for log-logistic distribution (95\% confidence level).}
  \begin{center}
    \begin{tabular}{| c  |  c  | c | c | c  |  c | c  |  c | c |} \hline
     $x_{L}$ [mm] & $N$     &  $\hat{\alpha}$  [mm]  & $\hat{\beta}$  & $\ln{L}$ & KS $(\sqrt{N}D)$    & AD $(A^2)$ &  KS & AD\\ \hline
  0    &  85     & $771.5$   & $11.6$  &-526.58  & 0.5549  & 0.6757 & $\checkmark$ & \ding{55} \\ 
\hline
  300 & 85     & $771.5$   & $11.6$  & -526.58 & 0.5551  & 0.6768 & $\checkmark$ & \ding{55}  \\ 
\hline
 400 & 84     & $774.1$   & $12.3$  & -514.65 & 0.5456  & 0.6626 & $\checkmark$ & $\checkmark$  \\ 
\hline
  500 & 83   &  $775.9$   & $12.7$  & -504.75 & 0.5803  & 0.5240 & $\checkmark$ & $\checkmark$ \\  \hline
      \end{tabular}%
  \end{center}
  \label{tab:precipitation_Toronto}
\end{table}%

\subsection{Outlook}
The log-logistic distribution is heavier tailed than the log-normal distribution and enjoys a closeness to the Pareto distribution. For modelling the distribution of wealth and income a left-truncated version of the log-logistic distribution~\cite{Fisk1961} is a sensible approach, because often very low incomes (below a certain threshold) in many economies are earned in a shadow economy. Therefore they are not captured in official government statistics. We have shown that the log-logistic hypothesis can explain certain cancer survival data very well and accurately, even passing KS and AD hypothesis tests. We can also confirm that in general the log-logistic distribution describes precipitation data regardless of their region fairly well. It is noteworthy that even in recent daily precipitation data there are days missing resulting in  a too ``small'' annual precipitation value if the missing data were replaced by zero. While the effect of this might be neglegible for higher annual precipitation it will have an effect for lower annual precipitation. A remedy for the statistical analysis could be the introduction of a left-truncation point to overcome this problem: we have seen that in this case both KS and AD tests are passed.

In recent literature it was suggested that one might investigate various modifications of the log-logistic distributions, such as, e.g., the Burr XII family of distributions \cite{Wingo1983}-\cite{Wang2010}.

Also in the case of mixed distributions where one mixture component is light tailed and the other is heavier tailed, such as log-logistic, our method will allow us to estimate the parameters of the log-logistic component with a good degree of accuracy. Thus, one has a tool at hand to decide whether or not a two-component mixture contains a log-logistic component.

\section{Proofs for theorems from Chapter 2}{\label{sec:proofs}}
\subsection{Proof of Theorem \ref{thm:exist}.}
We first begin with a lemma dealing with ``small'' $\lambda$ for the objective function given in Eq.~(\ref{eq:objective}). Keep in mind that we have chosen $x_L=1$.

\begin{mylemma}{\label{thm:asympt}}
For any $\beta>0$, there exists some $\lambda_0>0$ such that for $\lambda \in [0,\lambda_0)$ we have
\begin{eqnarray}
\varphi(\lambda,\beta) = \varphi(0,\beta) + N\left\{ -\frac{2}{N}\sum_{i=1}^N \frac{1}{X_i^{\beta}} +1 \right\} \lambda + {\cal{O}}(\lambda^2)
\label{eq:lambda_expansion}
\end{eqnarray}
where
\begin{eqnarray}
\varphi(0,\beta) = \lim_{\lambda\rightarrow 0+} \varphi(\lambda,\beta) = \sum_{i=1}^{N} \left[  -\beta\ln{X_i} +\ln{\beta} \right]
\label{eq:varphi_lambda_is_0}\quad.
\end{eqnarray}
\end{mylemma}

 {\em Proof:}

For fixed $\beta>0$ we obtain by differentiating Eq.~(\ref{eq:objective}) with respect to $\lambda$
\begin{eqnarray}
\frac{\partial}{\partial \lambda} ~\varphi(\lambda,\beta) & = & 
-\frac{N}{\lambda} \frac{ \frac{1}{\lambda}}{1+\frac{1}{\lambda}} - \frac{N}{\lambda}
+ \frac{2}{\lambda} \sum_{i=1}^N \frac{ \frac{X_i^{\beta}}{\lambda}}{1+\frac{X_i^{\beta}}{\lambda}}
\nonumber \\
& = & \frac{N}{\lambda} \left\{   \frac{2}{N} \sum_{i=1}^N   \frac{1}{1+\frac{\lambda}{X_i^{\beta}}}  -1 -  \frac{1}{1+\lambda}       \right\}
\nonumber \\
& = &  \frac{N}{\lambda } \left\{   -\frac{2}{N} \sum_{i=1}^N   \frac{1}{X_i^{\beta}}~\lambda  +\lambda +{\cal{O}}(\lambda^2)       \right\} \, ,
\label{eq:phi_prime}
\end{eqnarray}
where for $\lambda\in(0,\lambda_0)$ with $0<\lambda_0<1$ the geometric series is absolutely convergent. Thus integrating Eq.~(\ref{eq:phi_prime}) over $\lambda'$ from $0$ to $\lambda$ we get
\begin{eqnarray}
\varphi(\lambda,\beta) - \varphi(0,\beta) &= & \lim_{\varepsilon\rightarrow 0}\int_{\varepsilon}^{\lambda} d\lambda' ~N \left\{   -\frac{2}{N} \sum_{i=1}^N   \frac{1}{X_i^{\beta}}  +1 +{\cal{O}}(\lambda')       \right\}
\nonumber \\
& = & N\lambda \left\{   -\frac{2}{N} \sum_{i=1}^N   \frac{1}{X_i^{\beta}}  +1    \right\} +{\cal{O}}(\lambda^2)   \, . 
\nonumber
\end{eqnarray}
The computation of $\varphi(0,\beta)=\lim_{\varepsilon\rightarrow 0} \varphi(\varepsilon,\beta)$ is straight forward. $\blacksquare$

\begin{figure}[hptb]
	\begin{center}
                  \includegraphics[width=0.75\textwidth]{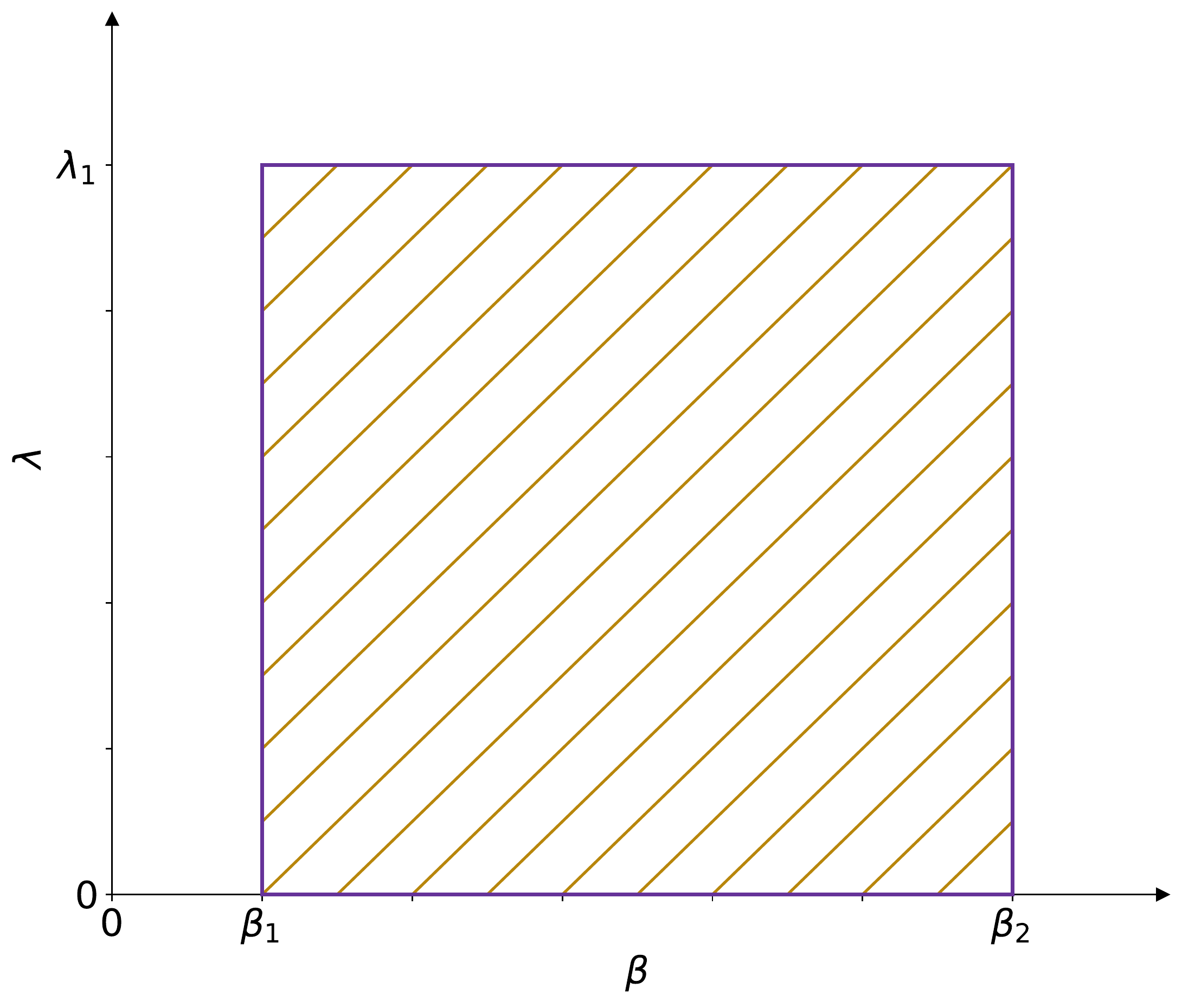}
		\caption{The closed intervall $[\beta_1,\beta_2]$ for the maximum of $\varphi(\lambda,\beta)$.\label{fig:region0}}
	\end{center}
\end{figure}

In the next lemma we derive an anologous result like in \cite{Gupta1999} about the behaviour of the objective function $\varphi(\lambda,\beta)$ in a certain rectangle in the first quadrant of the $(\lambda,\beta)$-coordinate plane. The situation is depicted in Fig. \ref{fig:region0}.

\begin{mylemma}{\label{thm:bounds}}
Consider an i.i.d. left-truncated sample $X_1,..., X_N$ all bigger than one and for which at least two observations are different. For any $M>0$ we have the following inequalities valid for 
$\varphi(\lambda,\beta)$ defined in Eq.~(\ref{eq:objective})

\begin{enumerate}[{(1)}]

\item There exists $\beta_1>0$ such that for $\beta \in(0,\beta_1)$ we have $\varphi(\lambda,\beta)<-M$, independent of $\lambda$

\item There exists $\beta_2>\beta_1$ such that for $\beta >\beta_2$ we have $\varphi(\lambda,\beta)<-M$, independent of $\lambda$

\item For $\beta\in [\beta_1,\beta_2]$ there exists $\lambda_1>0$ such that for $\lambda>\lambda_1$ we have also $\varphi(\lambda,\beta)<-M$

\end{enumerate}

\end{mylemma}

 {\em Proof:}

(1): We write Eq.~(\ref{eq:objective}) as
\begin{eqnarray}
\varphi(\lambda,\beta)
& = &  N \ln{ \left(1+\frac{1}{\lambda}  \right)} + N \ln{\beta} - N \ln{\lambda} + \beta S
\nonumber \\
& & 
-2 \sum_{i=1}^{N}  \ln{ \left[ 1+ \frac{X_i^{\beta}}{\lambda}   \right] }
\nonumber \\
& = & \sum_{i=1}^N  \ln{ \frac{1+\frac{1}{\lambda}  }{1+\frac{X_i^{\beta}}{\lambda}  } } + N \ln{\beta}  + \beta S 
+\sum_{i=1}^N  \ln{  \frac{\frac{1}{\lambda}  }{1+\frac{X_i^{\beta}}{\lambda}  } } 
\nonumber \\
& = & \sum_{i=1}^N  \ln{ \frac{\lambda+1}{\lambda+X_i^{\beta} } } + N \ln{\beta}  + \beta S 
+\sum_{i=1}^N  \ln{  \frac{1}{\lambda+X_i^{\beta}  } } 
\nonumber \\
& < & N \ln{\beta}  + \beta S
\label{eq:ineq1}
\end{eqnarray}
because $\lambda+X_i^{\beta}> \lambda + 1>1$. From the bound Eq.~(\ref{eq:ineq1}) we see that $\beta_1>0$ can be chosen independent of $\lambda\geq 0$ and the first statement is proven.

(2):  see proof in Appendix \ref{sec:ineq2}

(3): For $\lambda>1$ and $\beta\in[\beta_1,\beta_2]$ we have
\begin{eqnarray}
\varphi(\lambda,\beta)
 \leq   N \ln{ \left(1+1  \right)} + N \ln{\beta_2} - N \ln{\lambda} + \beta_2 S
\label{eq:ineq3}
\end{eqnarray}
because the last term in  Eq.~(\ref{eq:objective}) is always negative and the argument of the logarithm is always bigger than 1. From inequality Eq.~(\ref{eq:ineq3}) we see the existance of a sufficiently large $\lambda_1>1$ such that the third statement follows.
 $\blacksquare$

Now we have all the ingredients for the proof of Theorem~\ref{thm:exist}.
\\

 {\bf{Proof of Theorem \ref{thm:exist}.:}}

We start the proof by looking at Fig.~\ref{fig:region0}.

For statement (1) apply Lemma \ref{thm:bounds} to conclude that outside the shaded area our objective function $\varphi(\lambda,\beta)<-M$ for some arbitrarily large $M>0$, i.e. our objective function gets arbitrarily small. Thus a global maximum of the twice differentiable objective function 
$\varphi(\cdot,\cdot)$ is either inside the shaded region $[0,\lambda_1]\times[\beta_1,\beta_2]$ or on the boundary with $\lambda=0$, somewhere in the closed interval $[\beta_1,\beta_2]$. One easily computes the only critical point and thus a candidate for a maximum of the objective function on this boundary-line to be $\beta_0$, as given in Eq.~(\ref{eq:beta0}). By condition $\beta_0>\beta_C$ and Lemma \ref{thm:asympt} this can be excluded: 
there exists some small  $\lambda_0>0$ such that for $\lambda\in(0,\lambda_0)$
\begin{eqnarray}
\varphi(\lambda,\beta_0)>\varphi(0,\beta_0)
\nonumber
\end{eqnarray}
and statement (1) is proven. 

For statement (2) we use the following argument: by continuity there exists $\epsilon>0$ (first line) and by Lemma \ref{thm:asympt} a positive $\lambda_0$ (second line) such that
\begin{eqnarray}
\varphi(0,\beta_0) & = & \max{ \left\{\varphi(0,\beta)  | \beta \in [\beta_0-\epsilon, \beta_0+\epsilon] \right\}}
\nonumber \\
& \geq & \max{ \left\{ \varphi(\lambda,\beta)  | \beta \in [\beta_0-\epsilon, \beta_0+\epsilon] , \lambda \in [0,\lambda_0] \right\} }
\nonumber
\end{eqnarray}
and thus $\beta_0$ is local maximum.
$\blacksquare$

\subsection{Proof of Theorem  \ref{thm:profile}. }

For $\lambda>0$ we start from the rewritten Eq.~(\ref{eq:lambda}) 
\begin{eqnarray}
0 & = & \frac{2}{N}  \sum_{i=1}^{N}  \frac{ X_i^{\beta}}{\lambda+X_i^{\beta}}   -1-
\frac{ 1}{\lambda+1}
\label{eq:lambda_new}
\end{eqnarray}
and investigate this equation Eq.~(\ref{eq:lambda_new}) in greater detail. We firstly note the following

\begin{mylemma}{\label{thm:Lambda}}
Consider a left-truncated log-logistic sample $X_1,...,X_N$ all bigger than $1$. Define
the function $q:  \mathbb{R}_+ \rightarrow (0,1)$ by
\begin{eqnarray}
q\left(\beta|(X_1,...,X_N )\right)\equiv\frac{1}{N}\sum_{i=1}^{N} \left( \frac{1}{X_i} \right)^{\beta}
\nonumber
\end{eqnarray}
Then the following holds true

\begin{enumerate}[{(1)}]

\item $q(\cdot)$ is a monotonically decreasing function.

\item The equation $q\left(\beta|(X_1,...,X_N )\right)=\frac{1}{2}$ has a unique positive solution $\beta_C$.

\item Equation Eq.~(\ref{eq:lambda_new}) 
has for $\beta\in(0,\beta_C]$ only the trivial solution $\hat{\lambda}=\Lambda(\beta)=0$ and for $\beta>\beta_C$ in addition exactly one positive solution $\hat{\lambda}=\Lambda(\beta)>0$.

\item For any fixed $\beta >0 $ all the roots $\hat{\lambda}=\hat{\lambda}(\beta)$ of the MLE equation Eq  (\ref{eq:lambda_new}) 
depend continously on the sample $(X_1,...,X_N)$.

\item For  a fixed sample $(X_1,...,X_N)$  all the roots $\hat{\lambda}$ of the MLE equation Eq  (\ref{eq:lambda_new}) 
depend continously on $\beta \in (0,+\infty)$. In addition we have $\lim_{\beta\rightarrow \beta_C+}\Lambda(\beta)=0$, where $\Lambda(\beta)$ is defined in part (3) of the Lemma. 

\end{enumerate}

\end{mylemma}

{\bf{Proof}:}

(1) Note that $0<1/X_i<1...$ and therefore $(1/X_i)^{\beta}$ is monotonically decreasing and the first statement follows. 

(2) Next note that because of
\begin{eqnarray}
\lim_{\beta\rightarrow 0+} q\left(\beta|(X_1,...,X_N )\right) &=& 1
\nonumber
\\ 
\lim_{\beta\rightarrow \infty} q\left(\beta|(X_1,...,X_N )\right) &=& 0
\nonumber
\end{eqnarray}
and the monotonicity of  $q(\cdot)$ the second statment follows immediately.

(3) To prove the third statement we start from the likelihood equation, Eq.~(\ref{eq:lambda_new}),  defining for fixed $\beta>0$ for convenience   $Y_i=X_i^{\beta}$,  
\begin{eqnarray}
1+\frac{1}{\lambda+1} = \frac{2}{N} \sum_{i=1}^N \frac{Y_i}{\lambda+Y_i} \, .
\label{eq:lambda2}
\end{eqnarray}
Next define by the right-hand side of Eq.~(\ref{eq:lambda2}), the function 
\begin{eqnarray}
k(\lambda)=\frac{2}{N}  \sum_{i=1}^{N}  \frac{Y_i}{\lambda+Y_i}
\label{eq:k_funct}
\end{eqnarray}
which is monotonically decreasing in $\lambda \in [0,\infty)$, because
\begin{eqnarray}
k'(\lambda)=-\frac{2}{N}  \sum_{i=1}^{N}  \frac{Y_i}{(\lambda+Y_i)^2} < 0 \, .
\label{eq:k_slope}
\end{eqnarray}
Afterwards define a  function  by the left-hand side of  Eq.~(\ref{eq:lambda2}) as
\begin{eqnarray}
\ell(\lambda)=1+  \frac{1}{\lambda+1}
\label{eq:l_funct}
\end{eqnarray}
which is also monotonically decreasing in $\lambda \in [0,\infty)$, because
\begin{eqnarray}
\ell'(\lambda)= -\frac{1}{(\lambda+1)^2} < 0 \, .
\label{eq:l_slope}
\end{eqnarray}
Note that we have
\begin{eqnarray}
\lim_{\lambda\rightarrow 0+} k(\lambda) & = & 2
\nonumber \\
\lim_{\lambda \rightarrow +\infty} k(\lambda) & = & 0
\nonumber
\end{eqnarray}
and also  
\begin{eqnarray}
\lim_{\lambda\rightarrow 0+} \ell(\lambda) & = & 2
\nonumber \\
\lim_{\lambda\rightarrow +\infty} \ell(\lambda) & = & 1\quad.
\nonumber
\end{eqnarray}
Thus the trivial solution $\hat{\lambda}=0$ is always a solution. We next consider three cases:

{\bf Case I:} $k'(0)<\ell'(0)$

In this case according to Eq.~(\ref{eq:k_slope}) and Eq.~(\ref{eq:l_slope}) the function $k(\cdot)$ falls faster than the function $\ell(\cdot)$, at least for argument $\lambda$ near 0. If this is true for all $\lambda>0$, i.e. $k'(\lambda)<\ell'(\lambda)$, then clearly there will not be any intersection and 
$\hat{\lambda}=0$ is the only solution of Eq.~(\ref{eq:lambda_new}). If the latter is not true, then the situation will be as depicted in Fig. \ref{fig:function_graph_Case1}:
\begin{figure}[H]
	\begin{center}
                \includegraphics[width=0.7\textwidth]{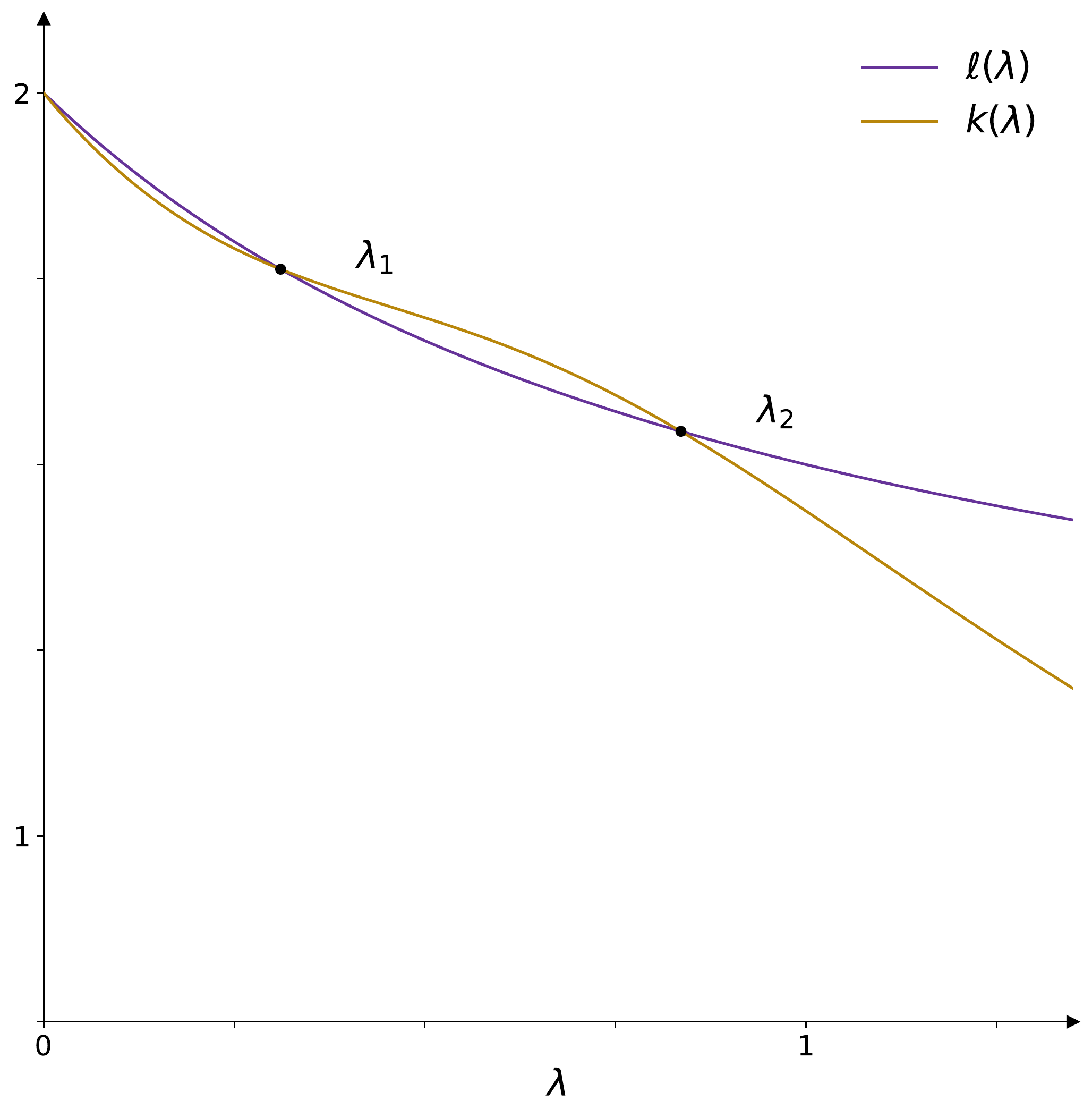}
		\caption{Case I: The functions $k(\lambda)$ (red) and $\ell(\lambda)$ (blue).\label{fig:function_graph_Case1}}
	\end{center}
\end{figure}
In this case $k(\cdot)$ initially falls quicker than $\ell(\cdot)$ and then somehow slows down to intersect at a point $\lambda_1>0$ and possibly a second time at a point $\lambda_2>\lambda_1$. Thus we have for $\lambda_1>0$ that the function graph of $k(\cdot)$ crosses the function graph of 
$\ell(\cdot)$ from below, meaning that the slopes satisfy the following inequality
\begin{eqnarray}
k'(\lambda_1) > \ell'(\lambda_1) \, .
\nonumber
\end{eqnarray}
By the intermediate value theorem there will exist a positive number $\lambda^* \in (0,\lambda_1)$ such that
\begin{eqnarray}
k'(\lambda^*) = \ell'(\lambda^*) \, .
\label{eq:contra}
\end{eqnarray}
We shall show that this will lead to a contradiction as follows. We write the Eq.~(\ref{eq:contra}) in detail as
\begin{eqnarray}
-\frac{2}{N}  \sum_{i=1}^{N}  \frac{Y_i}{(\lambda^*+Y_i)^2} = -\frac{1}{(\lambda^*+1)^2}
\nonumber
\end{eqnarray}
or rearranging
\begin{eqnarray}
\frac{1}{N}  \sum_{i=1}^{N}  {Y_i}\frac{(\lambda^*+1)^2}{(\lambda^*+Y_i)^2} = \frac{1}{2}\quad.
\label{eq:condition_case1}
\end{eqnarray}
Define the function $h(\cdot): \mathbb{R_+}\rightarrow  \mathbb{R_+}$ by
\begin{eqnarray}
h(\lambda) = \frac{1}{N}  \sum_{i=1}^{N}  {Y_i}\frac{(\lambda+1)^2}{(\lambda+Y_i)^2} \quad.
\label{eq:h_function}
\end{eqnarray}
Thus  Eq.~(\ref{eq:condition_case1}) is equivalent to   $h(\lambda^*)=\frac{1}{2}$. Note that 
\begin{eqnarray}
h'(\lambda) =\frac{1}{N}  \sum_{i=1}^{N}  {Y_i}\frac{2(\lambda+1)(\lambda+Y_i)-2(\lambda+1)^2}{(\lambda+Y_i)^3}  >0
\end{eqnarray}
because all terms are positive. In particular, the numerator is positive because $Y_i>1$. Note that we have the following limit 
\begin{eqnarray}
\lim_{\lambda\rightarrow\infty}h(\lambda) &=& \frac{1}{N}\sum_{i=1}^{N}  {Y_i} >1
\nonumber \\
\lim_{\lambda\rightarrow0+}h(\lambda) &=& \frac{1}{N}\sum_{i=1}^{N}  \frac{1}{Y_i} <1 \, .
\nonumber
\end{eqnarray}
Hence for the equation $h(\lambda)=\frac{1}{2}$ to have a solution in $(0,\lambda_1)$ we need to require that the second limit is smaller than $\frac{1}{2}$, i.e.
\begin{eqnarray}
\frac{1}{N}\sum_{i=1}^{N}  \frac{1}{Y_i} < \frac{1}{2}  \iff  k'(0) > \ell'(0) \, ,
\nonumber
\end{eqnarray}
which is a contradiction. Thus in Case I we have as the only real solution $\hat{\lambda}=\lambda_0=0$. Note that Case I corresponds to $\beta<\beta_C$.

{\bf Case II:} $k'(0)>\ell'(0)$ 

Here the function $k(\cdot)$  decreases slower than the function $\ell(\cdot)$ for $\lambda$ near 0 and only here can the graph of $k(\cdot)$ intersect the graph of $\ell(\cdot)$ from above at 
$\lambda_1>0$ before tending to zero as $\lambda$ gets larger, as depicted in 
Fig.~\ref{fig:function_graph_Case2}:
\begin{figure}[hptb]
	\begin{center}
                \includegraphics[width=0.7\textwidth]{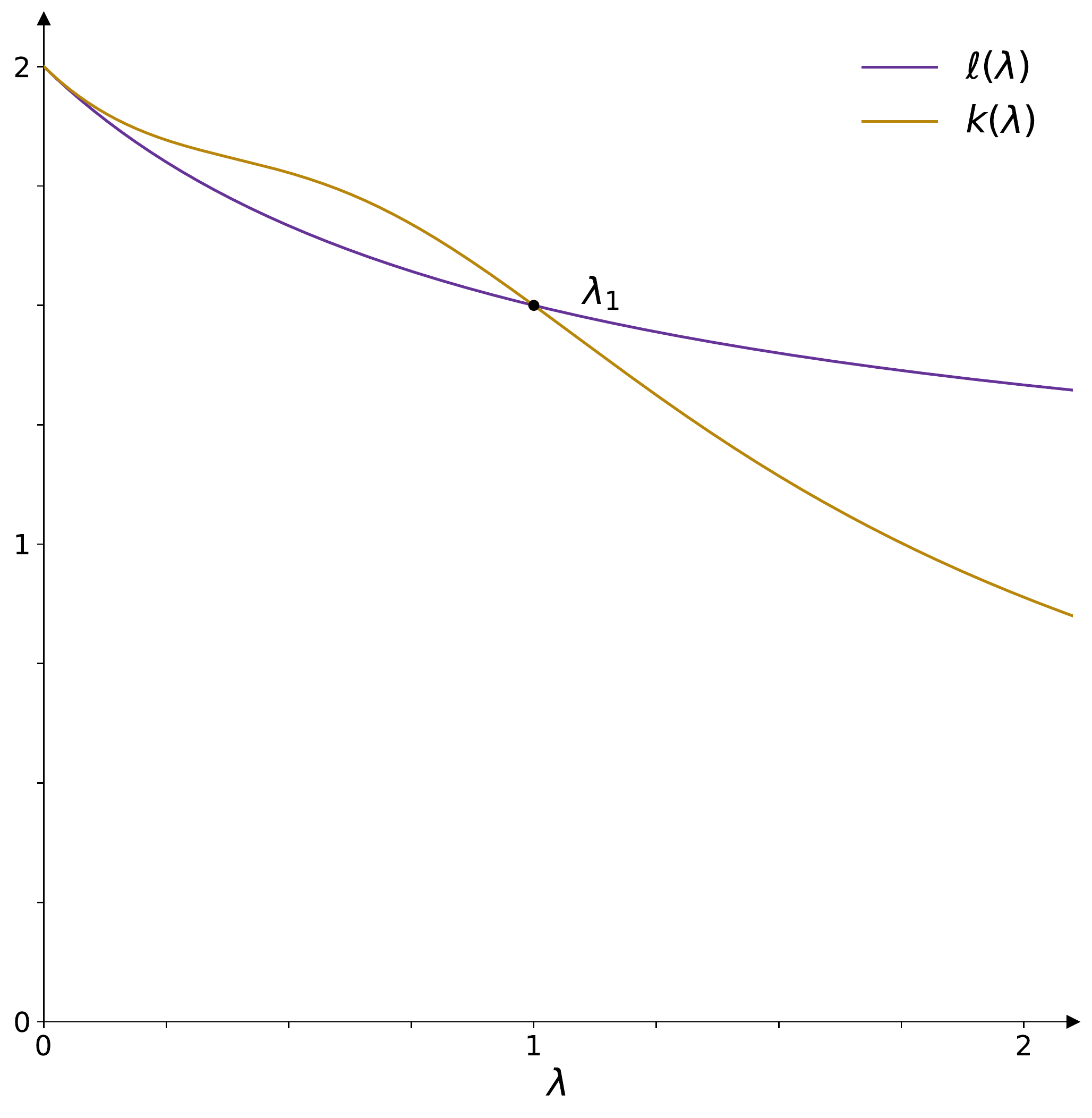}
		\caption{Case II: the functions $k(\lambda)$ (red) and $\ell(\lambda)$ (blue) \, .\label{fig:function_graph_Case2}}
	\end{center}
\end{figure}
We need to show now that this intersection is unique. We note that at $\lambda_1$ the graph of $k(\cdot)$ crosses the graph of $\ell(\cdot)$ from above, i.e.
\begin{eqnarray}
k'(\lambda_1)<\ell'(\lambda_1)\quad.
\label{eq:CaseII}
\end{eqnarray}
However, if there were a second crossing point, $\lambda_2>\lambda_1$, the graph of $k(\cdot)$ 
would cross the graph of $\ell(\cdot)$ from below, i.e.
\begin{eqnarray}
k'(\lambda_2)>\ell'(\lambda_2) \, .
\nonumber
\end{eqnarray}
As in Case I we conclude by the intermediate value theorem that there exists $\lambda^*\in (\lambda_1,\lambda_2)$ such that
\begin{eqnarray}
k'(\lambda^*)=\ell'(\lambda^*)
\nonumber
\end{eqnarray}
which is equivalent to 
\begin{eqnarray}
\frac{1}{N}  \sum_{i=1}^{N}  {Y_i}\frac{(\lambda^*+1)^2}{(\lambda^*+Y_i)^2} = \frac{1}{2}
\label{eq:condition_case2}
\end{eqnarray}
and this in turn is similar to Eq.~(\ref{eq:condition_case1}). We use the function $h(\cdot)$,
 as defined in Eq.~ (\ref{eq:h_function}), to obtain the following limits
\begin{eqnarray}
\lim_{\lambda\rightarrow +\infty} h(\lambda) &=& \frac{1}{N}\sum_{i=1}^N {Y_i }>1
\nonumber
\\
\lim_{\lambda\rightarrow \lambda_1} h(\lambda)&>& \frac{1}{2}
\nonumber
\end{eqnarray}
and Eq.~(\ref{eq:condition_case2}) has no solution. The last limit follows from Eq.~ (\ref{eq:CaseII}) and the definition of $h(\cdot)$ from Eq.~(\ref{eq:condition_case2})
\begin{eqnarray}
-\frac{2}{N} \sum_{i=1}^N \frac{Y_i}{(\lambda_1+Y_i)^2} < -\frac{1}{(\lambda_1+1)^2} \iff k'(\lambda_1)<\ell'(\lambda_1) \, .
\nonumber
\end{eqnarray}
Thus there exists only in this case one positive root $\lambda_1$ to  Eq.~(\ref{eq:lambda_new}). Case II corresponds to $\beta>\beta_C$.

{\bf Case III:} $k'(0)=\ell'(0)$ 

Here one needs to look at the next higher derivatives, i.e.  $k''(0) > \ell''(0)$, which reads here as
$ 1<\frac{2}{N}\sum_{i=1}^N \frac{1}{Y_i^2} $ or
\begin{eqnarray}
\frac{1}{2} < \frac{1}{N} \sum_{i=1}^N \left(\frac{1}{Y_i}\right)^2= q(2\beta) \, .
\nonumber
\end{eqnarray}
This cannot be true if
\begin{eqnarray}
\frac{1}{2} = \frac{1}{N} \sum_{i=1}^N \left(\frac{1}{Y_i}\right) = q(\beta) \, ,
\nonumber
\end{eqnarray}
because of the properties of the above defined function $q(\cdot)$. Note that Case III corresponds to 
$\beta=\beta_C$.Thus the third statement has been proven. 

(4) For the forth statement rewrite the MLE equation Eq.~(\ref{eq:lambda_new}) as in Eq.~(\ref{eq:lambda2})  and multiply both sides of the equation with $(\lambda+1)\prod_{i=1}^N (\lambda+Y_i)$. Thus 
after rearranging one obtains a polynomial of $(N+1)$-st order in the variable $\lambda$
\begin{eqnarray}
\lambda^{N+1}+b_N(Y_1,...,Y_N) \lambda^N+...+ b_0(Y_1,...,Y_N)=0
\end{eqnarray}
where the coefficients $b_i(Y_1,...Y_N)$ are certain continous polynomial functions in $(Y_1,...Y_N)$ tediously obtained in the process of re-arranging terms and sorting by the powers of $\lambda$. From Appendix~A in Ref.~\cite{Ostrowski1960}, the roots $\hat{\lambda}$ of some algebraic equation (i.e. a polynomial equation) depend continously on the coefficients and the statement is proven. 

(5) The last statement is shown in a similar manner to the previous statement and because all $X_i^{\beta}$ are continous functions in $\beta$ respectively, the desired statement is proven because for $\beta\leq\beta_C$ the only real root is $\hat{\lambda}=0$.
$\blacksquare$

The next Lemma provides some additional information about the function $\Lambda(\cdot)$ defined implicitly by Eq.~(\ref{eq:lambda_new}) via Lemma~\ref{thm:Lambda}. (3).

\begin{mylemma}
Consider a left-truncated log-logistic sample $X_1,...,X_N$ all bigger than $1$.  Then the non-zero solution of the MLE equation Eq.~(\ref{eq:lambda_new}) 
$\Lambda(\cdot): (\beta_C,\infty)\rightarrow \mathbb{R^+}$ has the following properties:

\begin{enumerate}[{(1)}]

\item For fixed $\beta$ $\Lambda(\beta)$ is a root of Eq.~(\ref{eq:lambda_new}) with multiplicity 1.

\item $\Lambda(\cdot)$ is an analytic function in the range of $(\beta_C,\infty)$.

\end{enumerate}

\end{mylemma}

{\bf Proof}:

Recall that  $x_L=1$.  
From Eq.~(\ref{eq:ML2}) respectively Eq.~(\ref{eq:objective}) we obtain for fixed $\beta\in(\beta_C,+\infty)$ by direct computation
\begin{eqnarray}
\left. \frac{\partial}{\partial\lambda}~\varphi(\lambda,\beta) \right|_{\Lambda(\beta)} &=&\frac{N}{\Lambda(\beta)} \left[ \frac{2}{N} \sum_{i=1}^N \frac{X_i^{\beta}}{\lambda+X_i^{\beta}}   -  1-\frac{1}{\lambda+1}\right]_{\Lambda(\beta)}=0
\nonumber \\
\left. \frac{\partial^2}{\partial\lambda^2}~\varphi(\lambda,\beta) \right|_{\Lambda(\beta)} &=&
-\frac{N}{\Lambda(\beta)} \left[ \frac{2}{N} \sum_{i=1}^N \frac{X_i^{\beta}}{(\lambda+X_i^{\beta})^2}   -\frac{1}{(\lambda+1)^2}\right]_{\Lambda(\beta)}
\nonumber
\end{eqnarray}
where the first derivative is equal to zero for $\lambda=\Lambda(\beta)$ by construction. Next we define a polynomial function $p(\cdot):\mathbb{R}^{+}_{0} \rightarrow \mathbb{R}$ by
\begin{eqnarray}
p(\lambda) &\equiv& \left[ \frac{2}{N} \sum_{i=1}^N \frac{X_i^{\beta}}{\lambda+X_i^{\beta}}-\frac{\lambda+2}{\lambda+1}\right]\, (\lambda+1) \prod_{j=1}^N (\lambda+X_j^{\beta}) 
\nonumber \\
& = & \frac{2}{N} \sum_{i=1}^N X_i^{\beta} (\lambda+1)\prod_{1\le j\le N\atop j\ne i} (\lambda+X_j^{\beta})
- (\lambda+2)\prod_{j=1}^N (\lambda+X_j^{\beta}) 
\label{eq:aux_funct_poly}
\end{eqnarray}
The first claim of the Lemma states that $\Lambda(\beta)$ is a root of $p(\cdot)$, Eq.~(\ref{eq:aux_funct_poly}), but not of $p'(\cdot)$. By construction we have indeed  $p(\Lambda(\beta))=0$. We show that the assumption $p'(\Lambda(\beta))=0$ leads to a contradiction. Compute the derivative in Eq.~(\ref{eq:aux_funct_poly})
\begin{eqnarray}
p'(\lambda) &=&  \frac{2}{N} \sum_{i=1}^N X_i^{\beta} \prod_{1\le j\le N\atop j\ne i} (\lambda+X_j^{\beta}) 
+ \frac{2}{N} \sum_{i=1}^N X_i^{\beta} (\lambda+1) \sum_{k=1}^N \prod_{1\le j\le N\atop j\ne i, j\ne k} (\lambda+X_j^{\beta}) 
\nonumber \\
& &
- \prod_{j=1}^N (\lambda+X_j^{\beta})- (\lambda+2) \sum_{k=1}^N \prod_{1\le j \le N \atop  j\ne k} (\lambda+X_j^{\beta}) \quad.
\nonumber
\end{eqnarray}
Then write down $p'(\Lambda(\beta))/\prod_{j}(\Lambda(\beta)+X_j^{\beta})$, which  by our asumption $p'(\Lambda(\beta))=0$ vanishes
\begin{eqnarray}
0 &=& \frac{2}{N} \sum_{i=1}^N \frac{X_i^{\beta}}{\Lambda(\beta)+X_i^{\beta}} + \frac{2}{N} (\Lambda(\beta)+1) \sum_{i=1}^N \frac{X_i^{\beta}}{\Lambda(\beta)+X_i^{\beta}} \sum_{k=1}^N \frac{1}{\Lambda(\beta)+X_k^{\beta}}
\nonumber \\
& & -1 -(\Lambda(\beta)+2)\sum_{k=1}^N \frac{1}{\Lambda(\beta)+X_k^{\beta}}
\nonumber \\
& = & \frac{2}{N} \sum_{i=1}^N \frac{X_i^{\beta}}{\Lambda(\beta)+X_i^{\beta}} \left[ 1+(\Lambda(\beta)+1)\sum_{k=1}^N \frac{1}{\Lambda(\beta)+X_k^{\beta}} \right]
\nonumber \\
& & -1 -(\Lambda(\beta)+2)\sum_{k=1}^N \frac{1}{\Lambda(\beta)+X_k^{\beta}}
\nonumber \\
& = & \frac{\Lambda(\beta)+2}{\Lambda(\beta)+1} \left[ 1+(\Lambda(\beta)+1)\sum_{k=1}^N \frac{1}{\Lambda(\beta)+X_k^{\beta}} \right]
\nonumber \\
& & -1 -(\Lambda(\beta)+2)\sum_{k=1}^N \frac{1}{\Lambda(\beta)+X_k^{\beta}}
\nonumber \\
& = & \frac{\Lambda(\beta)+2}{\Lambda(\beta)+1} -1 = \frac{1}{\Lambda(\beta)+1}
\nonumber
\end{eqnarray}
which is a contradiction, because $\Lambda(\beta)>0$ and thus the first claim is proven.
Hence for $\beta \in (\beta_C,+\infty)$ fixed and any critical point $\Lambda(\beta)$ of $\varphi(\cdot,\beta)$ we have always
\begin{eqnarray}
\left. \frac{\partial^2}{\partial\lambda^2}\varphi(\lambda,\beta)\right|_{\Lambda(\beta)} &\neq& 0
\nonumber
\end{eqnarray}
We know that for a maximum the second derivative is negative, so we have always
\begin{eqnarray}
\left. \frac{\partial^2}{\partial\lambda^2}\varphi(\lambda,\beta) \right|_{\Lambda(\beta)} &<& 0
\nonumber
\end{eqnarray}
from which we conclude that for all $\beta>\beta_C$
\begin{eqnarray}
 \left[ \frac{2}{N} \sum_{i=1}^N \frac{X_i^{\beta}}{(\lambda+X_i^{\beta})^2}   -\frac{1}{(\lambda+1)^2}\right]_{\Lambda(\beta)}>0
\label{eq:secondorder}
\end{eqnarray}

For the second claim we note that by its definition as root of multiplicity 1 the function $\Lambda(\cdot)$ is analytic for argument $\beta>\beta_C$ by simple application of the implicit function theorem (e.g. \cite{Ostrowski1960})

and the proof is finished.
$\blacksquare$

Now we have all the ingredients for the proof of  Theorem~\ref{thm:profile}.
\\

 {\bf Proof of Theorem \ref{thm:profile}.:}

Lemma~\ref{thm:Lambda}. (3) contains already the main statements concerning the unique construction of the function $\Lambda(\cdot):\mathbb{R}_{0}^{+}\rightarrow  \mathbb{R}^{+}_{0}$ . The behaviour of the derivatives  $\frac{\partial}{\partial \lambda}\varphi(\lambda,\beta)$ for $\lambda \neq \Lambda(\beta)$ follows from the easily obtained asymptotics for $\lambda \rightarrow \infty$:
\begin{eqnarray}
\frac{\partial}{\partial \lambda}\varphi(\lambda,\beta) =-\frac{N}{\lambda} \left\{ 1+  +{\cal{O}}\left(\frac{1}{\lambda} \right)\right\}
\nonumber
\end{eqnarray}
and Lemma \ref{thm:asympt} for small $\lambda\geq 0$ and the fact that $\frac{\partial}{\partial \lambda}\varphi(\lambda,\beta)$ has only one zero.$\blacksquare$

The next Lemma treats the situation where the log-logistic distribution degenerates when $\lambda\rightarrow 0+$. We formulate it for the convencience of general $x_L>0$.
\begin{mylemma}{\label{thm:Pareto}}
Consider the left-truncated log-logistic distribution given in Eq.~(\ref{eq:pdf_trunc}) with $x_L>0$. In the limit $\lambda \rightarrow 0+$ respectively $\alpha \rightarrow 0+$ the following holds true
\begin{enumerate}[{(1)}]
\item There is $L^1$-convergence in the sense of Lebesgue of
\begin{eqnarray}
f_{LT}(\cdot|\alpha,\beta) \xrightarrow{\text{$L^1$}} g(\cdot|\beta)  \hspace{1cm} \text{as}  \hspace{0.5cm} \alpha\rightarrow 0+
\nonumber
\end{eqnarray}
where
\begin{eqnarray}
g(x|\beta)= \left( \frac{\beta}{x_L} \right) \left(\frac{x}{x_L}\right)^{-(1+\beta)} 
\label{eq:Pareto_pdf}
\end{eqnarray}
is the Pareto probability distribution density.

\item While the first ML equation for $\alpha$ respectively $\lambda$, Eq.~(\ref{eq:lambda}), becomes redunant, the second ML equation for $\beta$, Eq.~(\ref{eq:beta_new}) simplifies to
\begin{eqnarray}
\frac{1}{\beta} =  \frac{1}{N}\sum_{i=1}^{N} \ln{\left( \frac{X_i}{x_L} \right)}
\label{eq:beta_Pareto1}
\end{eqnarray}

\end{enumerate}

\end{mylemma}

{\bf Proof}:

(1) To prove statement (1) we first we show pointwise convergence for some $x>x_L$ fixed, namly if we chose any arbitrary positive sequence $(\alpha_n)_{n\in  \mathbb{N}} \rightarrow 0+$ as $n\rightarrow \infty$
\begin{eqnarray}
&& \left[1+ \left( \frac{x_L}{\alpha_n} \right)^{\beta}\right]  \frac{\beta}{\alpha_n} \left( \frac{x}{\alpha_n} \right)^{\beta-1}  \frac{1}{\left[1+\left(\frac{x}{\alpha_n}\right)^{\beta}\right]^2} 
\nonumber \\
& =  & \left( \frac{x_L}{\alpha_n} \right)^{\beta}  \frac{\beta}{\alpha_n} \left( \frac{x}{\alpha_n} \right)^{\beta-1}  \left( \frac{x}{\alpha_n} \right)^{-2\beta} + \hspace{0.25cm} \mathcal{O}(\alpha_n^{2\beta})
\nonumber \\
&  \xrightarrow{n \rightarrow \infty} & \left( \frac{\beta}{x_L} \right) \left(\frac{x}{x_L}\right)^{-\beta-1}  =  g(x|\beta)\quad.
\nonumber
\end{eqnarray}
Next we need to show that the functions $f_{LT}(\cdot|\alpha_n,\beta)$ can be bounded by a positive integrable function. We have decided for the following bound
\begin{eqnarray}
0 < f_{LT}(x|\alpha_n,\beta) < c \cdot g(x|\beta)
\nonumber
\end{eqnarray}
where the constant $c>0$ needs to be chosen in a suitable way. To this end we start from the following
\begin{eqnarray}
\left[1+ \left( \frac{x_L}{\alpha_n} \right)^{\beta}\right]  \frac{\beta}{\alpha_n} \left( \frac{x}{\alpha_n} \right)^{\beta-1}  \frac{1}{\left[1+\left(\frac{x}{\alpha_n}\right)^{\beta}\right]^2} & < & c \left( \frac{\beta}{x_L} \right) \left(\frac{x}{x_L}\right)^{-(1+\beta)}
\nonumber
\end{eqnarray}
which can be reduced to
\begin{eqnarray}
\frac{\alpha_n^{\beta}+x_L^{\beta}}{x_L^{\beta}} \left( \frac{x^{\beta}}{\alpha_n^{\beta}+x^{\beta}}    \right)^2 < c \, .
\nonumber
\end{eqnarray}
Now there exists $n_0>0$ such that $(\alpha_n^{\beta}+x_L^{\beta})/x_L^{\beta}\leq 2$ for $n>n_0$. Because $x^{\beta}/(\alpha_n^{\beta}+x^{\beta})<1$ we may chose
\begin{eqnarray}
c = \max{ \left\{2, \max_{0\leq n \leq n_0} \left( 1+\frac{\alpha_n^{\beta}}{x_L^{\beta}} \right) \right\} }
\nonumber
\end{eqnarray}
The desired convergence result in $L^1$-Lebesgue sense follows from the Lebesgue dominated convergence theorem.

(2) To prove statement (2) note that in Eq.~(\ref{eq:beta_new}) as $\alpha_n \rightarrow 0+$ the leading term on the right-hand side is $1/\beta + \ln{\left(x_L/\alpha_n\right)}$ whereas the leading term on the left-hand side is $\sum_{i=1}^N \ln{\left(X_i/\alpha_n\right)}\cdot 1$ and re-arranging terms and performing the limit yields the desired result.
 $\blacksquare$

\subsection{Proof of Corollary \ref{thm:master}.}
{\bf Proof:}

Part (1) is an immediate consequence of our Theorems. The proof of statement (2) is provided by the Lemma~\ref{thm:Pareto}.
 $\blacksquare$

\subsection{Proof of Theorem \ref{thm:identical}.}
{\em Proof:}

(1): The expressions follow immediately from Theorem~\ref{thm:exist}. Note that $X_1>1$ throughout.

(2): Because $\lambda=\Lambda(\beta)>0$ for $\beta>\beta_C$ we can write Eq.~(\ref{eq:lambda}) as
\begin{eqnarray}
0 = 2 \frac{X_1^{\beta}}{\lambda+X_1^{\beta}} - 1 - \frac{1}{\lambda+1}
\nonumber
\end{eqnarray}
Now inserting $\lambda=\Lambda(\beta)=X_1^{\beta}-2$ on the right-hand side yields indeed
\begin{eqnarray}
2 \frac{X_1^{\beta}}{2( X_1^{\beta}-1)} - 1 - \frac{1}{X_1^{\beta}-1} = \frac{X_1^{\beta}}{ X_1^{\beta}-1} - \frac{X_1^{\beta}-1}{X_1^{\beta}-1}-\frac{1}{X_1^{\beta}-1} = 0 \, .
\nonumber
\end{eqnarray}

ad (3): We see continuity by straight-forward computation 
\begin{eqnarray}
\lim_{\beta\rightarrow \beta_C-} \tilde{\varphi}(\beta)= \lim_{\beta\rightarrow \beta_C+} \tilde{\varphi}(\beta) = N\ln{\beta_C}-N\ln{2}\quad.
\nonumber
\end{eqnarray}
For monotonicity we differentiate the profile likelihood with respect to $\beta$. For $\beta \in (0,\beta_C)$ we obtain
\begin{eqnarray}
\tilde{\varphi}'(\beta)=N \left(\frac{1}{\beta} - \ln{X_1} \right) >0
\nonumber
\end{eqnarray}
because $\beta<\beta_C=\ln{2}/\ln{X_1} < 1/\ln{X_1}$.
Finally for $\beta \in (\beta_C,+\infty)$ we obtain after some easy computation
\begin{eqnarray}
\tilde{\varphi}'(\beta)=N \left(\frac{1}{\beta} - \frac{\ln{X_1}}{X_1^{\beta}-1} \right) 
= \left(\frac{1}{\beta} - \frac{1}{\beta_C}\frac{\ln{2}}{X_1^{\beta}-1} \right) >0
\nonumber
\end{eqnarray}
where the last inequality follows from the inequality
\begin{eqnarray}
\beta_C (X_1^{\beta}-1)> \beta \ln{2} \, .
\nonumber
\end{eqnarray}
To see this define two functions $F(\beta) = \beta_C (X_1^{\beta}-1)$ and 
$G(\beta) = \beta \ln{2}$. Obviously we have $F(\beta_C)=\beta_C>G(\beta_C)=\beta_C \ln{2}$.
With the derivatives $F'(\beta)=\beta_C X_1^{\beta}\ln{X_1}=X_1^{\beta}\ln{2}>G'(\beta)=\ln{2}$ we obtain the desired result by
\begin{eqnarray}
F(\beta)-G(\beta)= F(\beta_C)-G(\beta_C)+ \int_{\beta_C}^{\beta} db~\left(F'(b)-G'(b) \right) >0
\nonumber
\end{eqnarray}
because all terms on the right-hand side are positive.
 $\blacksquare$

\newpage
\appendix
\section{Tables for $95 \%$ confidence levels}\label{sec:Tables_95_percentail}
    \leavevmode\vfill
    \begingroup
\vfill
\hfil
\centering
\begin{sideways}
 \begin{threeparttable}
\centering
  \caption{left-truncated log-logistic KS($\sqrt{N}D$) significance level 95\%}
 \begin{tabular}{|rr|ccccccc|}
\hline
    p    & $\sqrt{\eta}$ & 30 & 50 & 100 & 200 & 500 & 1000 & 10000 \\
\hline
    0     & 0     & 0.7661(06) & 0.7774(07) & 0.7860(07) & 0.7916(08) & 0.7976(07) & 0.8002(07) & 0.8039(07) \\
    0.0323 & 0.18  & 0.7602(06) & 0.7711(07) & 0.7799(07) & 0.7871(07) & 0.7915(06) & 0.7942(08) & 0.7981(06) \\
    0.1   & 0.33  & 0.7640(07) & 0.7746(07) & 0.7840(07) & 0.7898(07) & 0.7952(07) & 0.7982(07) & 0.8017(07) \\
    0.2   & 0.5   & 0.7705(06) & 0.7809(06) & 0.7912(07) & 0.7972(06) & 0.8017(07) & 0.8048(08) & 0.8095(08) \\
    0.3   & 0.65  & 0.7761(07) & 0.7869(07) & 0.7975(08) & 0.8032(07) & 0.8090(06) & 0.8117(07) & 0.8148(07) \\
    0.4   & 0.82  & 0.7795(07) & 0.7919(07) & 0.8024(08) & 0.8094(08) & 0.8144(07) & 0.8171(08) & 0.8211(07) \\
    0.5   & 1     & 0.7832(07) & 0.7952(08) & 0.8071(07) & 0.8138(07) & 0.8192(07) & 0.8210(07) & 0.8253(07) \\
    0.6   & 1.22  & 0.7843(06) & 0.7981(07) & 0.8098(07) & 0.8164(07) & 0.8224(07) & 0.8248(07) & 0.8292(07) \\
    0.7   & 1.53  & 0.7865(07) & 0.7991(07) & 0.8124(08) & 0.8205(07) & 0.8258(07) & 0.8287(08) & 0.8326(07) \\
    0.8   & 2     & 0.7880(07) & 0.8015(07) & 0.8146(08) & 0.8215(07) & 0.8292(07) & 0.8319(07) & 0.8352(08) \\
    0.8605 & 2.48  & 0.7885(07) & 0.8015(08) & 0.8144(08) & 0.8227(07) & 0.8301(07) & 0.8328(08) & 0.8375(07) \\
    0.9   & 3     & 0.7884(07) & 0.8028(07) & 0.8157(08) & 0.8242(08) & 0.8310(07) & 0.8339(08) & 0.8393(07) \\
\hline
    \end{tabular}%
  \label{tab:KS95}%
 \centering
\caption{left-truncated log-logistic AD($A^2$) significance level 95\% }
 \begin{tabular}{|rr|ccccccc|}
\hline
    p    & $\sqrt{\eta}$ & 30 & 50 & 100 & 200 & 500 & 1000 & 10000 \\
\hline
    0     & 0     & 0.6594(12) & 0.6614(14) & 0.6632(12) & 0.6634(15) & 0.6647(13) & 0.6653(14) & 0.6652(13) \\
    0.0323 & 0.18  & 0.6580(12) & 0.6589(14) & 0.6572(12) & 0.6598(13) & 0.6592(12) & 0.6595(13) & 0.6589(13) \\
    0.1   & 0.33  & 0.6695(13) & 0.6689(14) & 0.6707(13) & 0.6699(12) & 0.6699(13) & 0.6705(13) & 0.6698(13) \\
    0.2   & 0.5   & 0.6857(14) & 0.6845(14) & 0.6851(13) & 0.6855(13) & 0.6849(14) & 0.6865(14) & 0.6869(13) \\
    0.3   & 0.65  & 0.6972(12) & 0.6987(13) & 0.6995(14) & 0.6995(15) & 0.7006(13) & 0.6996(14) & 0.6996(15) \\
    0.4   & 0.82  & 0.7055(14) & 0.7091(15) & 0.7108(16) & 0.7113(13) & 0.7110(15) & 0.7114(15) & 0.7127(15) \\
    0.5   & 1     & 0.7133(14) & 0.7166(15) & 0.7203(14) & 0.7224(15) & 0.7217(13) & 0.7224(15) & 0.7222(16) \\
    0.6   & 1.22  & 0.7161(14) & 0.7235(16) & 0.7280(15) & 0.7296(14) & 0.7319(14) & 0.7312(14) & 0.7317(15) \\
    0.7   & 1.53  & 0.7219(14) & 0.7270(15) & 0.7340(15) & 0.7376(15) & 0.7392(15) & 0.7400(15) & 0.7388(14) \\
    0.8   & 2     & 0.7231(16) & 0.7316(16) & 0.7394(15) & 0.7403(15) & 0.7460(16) & 0.7479(15) & 0.7468(15) \\
    0.8605 & 2.48  & 0.7256(15) & 0.7314(16) & 0.7400(13) & 0.7440(16) & 0.7496(16) & 0.7513(18) & 0.7523(15) \\
    0.9   & 3     & 0.7252(17) & 0.7345(16) & 0.7399(17) & 0.7469(16) & 0.7507(16) & 0.7525(14) & 0.7541(16) \\
\hline
    \end{tabular}%
  \label{tab:AD95}%
  %
  %
  \caption{Interpolation formula Eq.(\ref{eq:interpol}): coefficients for different significance levels}
    \begin{tabular}{|r|r|rrrrrrrrr|}
    \hline
          & s.l. & $\theta_1$ & $\theta_2$ & $\theta_3$ & $\theta_4$ & $\theta_5$ & $\theta_6$ & $\theta_7$ & $\theta_8$ & $\theta_9$ \\
    \hline
    AD ($A^2$) & 85\%  & 0.5644 & -0.0026 & 0.1307 & 0.0406 & 0.2612 & -0.0432 & 0.0001 & 0.0350 & -0.1715 \\
    AD ($A^2$) & 90\%  & 0.6390 & -0.0005 & 0.1540 & 0.0361 & 0.2750 & -0.0497 & 0.0000 & 0.0398 & -0.2066 \\
    AD ($A^2$) & 95\%  & 0.7669 & -0.0189 & 0.1927 & 0.0094 & 0.2914 & -0.0641 & 0.0001 & 0.0494 & -0.2364 \\
    AD ($A^2$) & 99\%  & 1.0714 & -0.0589 & 0.2933 & 0.0301 & 0.3272 & -0.0943 & 0.0002 & 0.0598 & -0.1787 \\
    KS ($\sqrt{n}D$) & 85\%  & 0.7421 & -0.0492 & 0.1565 & -0.0517 & 0.2210 & -0.0238 & 0.0000 & 0.1492 & -0.2115 \\
    KS ($\sqrt{n}D$) & 90\%  & 0.7821 & -0.0818 & 0.1742 & -0.0917 & 0.2336 & -0.0298 & 0.0001 & 0.1500 & -0.2723 \\
    KS ($\sqrt{n}D$) & 95\%  & 0.8443 & -0.1204 & 0.1987 & -0.1318 & 0.2470 & -0.0298 & 0.0001 & 0.1417 & -0.3753 \\
    KS ($\sqrt{n}D$) & 99\%  & 0.9711 & -0.1729 & 0.2672 & -0.1687 & 0.2895 & -0.0392 & 0.0001 & 0.1416 & -0.5325 \\
    \hline
    \end{tabular}%
  \label{tab:theta}%
   \end{threeparttable}
\end{sideways}
   \vfill
\endgroup
%
\newpage
\section{Tables for other confidence levels}\label{sec:Tables}
    \leavevmode\vfill
    \begingroup
\vfill
\hfil
\centering
\small
\begin{sideways}
  \begin{threeparttable}
  \centering
  \caption{left-truncated log-logistic KS($\sqrt{N}D$) significance level 85\%}
 \begin{tabular}{|rr|ccccccc|}
\hline
    p    & $\sqrt{\eta}$ & 30 & 50 & 100 & 200 & 500 & 1000 & 10000 \\
\hline
    0     & 0     & 0.6740(04) & 0.6830(05) & 0.6907(05) & 0.6961(05) & 0.7012(04) & 0.7036(04) & 0.7072(05) \\
    0.0323 & 0.18  & 0.6687(05) & 0.6775(04) & 0.6854(04) & 0.6913(05) & 0.6961(04) & 0.6982(05) & 0.7019(04) \\
    0.1   & 0.33  & 0.6719(04) & 0.6806(05) & 0.6891(04) & 0.6941(05) & 0.6993(04) & 0.7022(04) & 0.7053(05) \\
    0.2   & 0.5   & 0.6774(04) & 0.6864(04) & 0.6943(04) & 0.7000(05) & 0.7047(05) & 0.7075(05) & 0.7117(05) \\
    0.3   & 0.65  & 0.6818(04) & 0.6909(05) & 0.6997(05) & 0.7056(05) & 0.7104(04) & 0.7130(04) & 0.7162(05) \\
    0.4   & 0.82  & 0.6846(04) & 0.6948(05) & 0.7040(06) & 0.7098(05) & 0.7145(05) & 0.7171(05) & 0.7208(05) \\
    0.5   & 1     & 0.6875(05) & 0.6974(05) & 0.7078(04) & 0.7139(05) & 0.7184(05) & 0.7207(05) & 0.7246(05) \\
    0.6   & 1.22  & 0.6889(05) & 0.7000(04) & 0.7098(05) & 0.7163(04) & 0.7218(05) & 0.7239(06) & 0.7278(04) \\
    0.7   & 1.53  & 0.6904(05) & 0.7011(05) & 0.7119(05) & 0.7189(05) & 0.7240(05) & 0.7269(05) & 0.7306(05) \\
    0.8   & 2     & 0.6916(05) & 0.7025(05) & 0.7139(04) & 0.7200(05) & 0.7268(05) & 0.7294(05) & 0.7331(05) \\
    0.8605 & 2.48  & 0.6920(04) & 0.7032(04) & 0.7140(05) & 0.7211(05) & 0.7275(04) & 0.7305(05) & 0.7344(05) \\
    0.9   & 3     & 0.6922(05) & 0.7038(04) & 0.7140(05) & 0.7220(05) & 0.7281(05) & 0.7309(05) & 0.7357(05) \\
\hline
    \end{tabular}%
  \label{tab:KS85}%
  \centering
  \caption{left-truncated log-logistic KS($\sqrt{N}D$) significance level 90\%}
 \begin{tabular}{|rr|ccccccc|}
\hline
    p    & $\sqrt{\eta}$ & 30 & 50 & 100 & 200 & 500 & 1000 & 10000 \\
\hline
    0     & 0     & 0.7101(05) & 0.7199(06) & 0.7280(05) & 0.7334(06) & 0.7388(05) & 0.7412(05) & 0.7450(06) \\
    0.0323 & 0.18  & 0.7047(05) & 0.7139(05) & 0.7225(05) & 0.7286(05) & 0.7334(05) & 0.7356(06) & 0.7394(05) \\
    0.1   & 0.33  & 0.7079(05) & 0.7174(05) & 0.7262(05) & 0.7316(06) & 0.7368(05) & 0.7398(05) & 0.7429(06) \\
    0.2   & 0.5   & 0.7139(05) & 0.7236(06) & 0.7322(05) & 0.7381(05) & 0.7428(06) & 0.7457(06) & 0.7496(05) \\
    0.3   & 0.65  & 0.7186(05) & 0.7285(05) & 0.7380(06) & 0.7441(06) & 0.7490(05) & 0.7514(05) & 0.7548(05) \\
    0.4   & 0.82  & 0.7219(05) & 0.7331(05) & 0.7425(06) & 0.7486(06) & 0.7540(06) & 0.7564(06) & 0.7600(06) \\
    0.5   & 1     & 0.7251(05) & 0.7357(05) & 0.7468(05) & 0.7531(06) & 0.7579(05) & 0.7604(06) & 0.7639(06) \\
    0.6   & 1.22  & 0.7266(05) & 0.7385(05) & 0.7489(06) & 0.7554(06) & 0.7611(06) & 0.7638(06) & 0.7674(05) \\
    0.7   & 1.53  & 0.7282(05) & 0.7397(06) & 0.7511(06) & 0.7586(06) & 0.7639(06) & 0.7669(06) & 0.7705(05) \\
    0.8   & 2     & 0.7295(06) & 0.7410(05) & 0.7533(05) & 0.7599(06) & 0.7670(05) & 0.7697(05) & 0.7734(05) \\
    0.8605 & 2.48  & 0.7300(05) & 0.7417(05) & 0.7532(05) & 0.7609(06) & 0.7676(06) & 0.7707(06) & 0.7747(05) \\
    0.9   & 3     & 0.7299(05) & 0.7424(05) & 0.7539(06) & 0.7622(06) & 0.7684(06) & 0.7714(05) & 0.7763(06) \\
\hline
    \end{tabular}%
  \label{tab:KS90}%
  \centering
   \caption{left-truncated log-logistic KS($\sqrt{N}D$) significance level 99\%}
 \begin{tabular}{|rr|ccccccc|}
\hline
    p    & $\sqrt{\eta}$ & 30 & 50 & 100 & 200 & 500 & 1000 & 10000 \\
\hline
    0     & 0     & 0.8780(11) & 0.8918(14) & 0.9015(15) & 0.9093(13) & 0.9160(14) & 0.9180(13) & 0.9227(14) \\
    0.0323 & 0.18  & 0.8715(12) & 0.8858(12) & 0.8958(14) & 0.9041(14) & 0.9100(13) & 0.9124(12) & 0.9171(14) \\
    0.1   & 0.33  & 0.8754(14) & 0.8884(14) & 0.9005(13) & 0.9078(13) & 0.9128(13) & 0.9165(13) & 0.9206(13) \\
    0.2   & 0.5   & 0.8829(11) & 0.8956(14) & 0.9081(12) & 0.9167(12) & 0.9216(12) & 0.9236(13) & 0.9292(14) \\
    0.3   & 0.65  & 0.8895(13) & 0.9037(14) & 0.9163(14) & 0.9224(13) & 0.9293(14) & 0.9326(13) & 0.9360(14) \\
    0.4   & 0.82  & 0.8933(14) & 0.9110(13) & 0.9225(14) & 0.9309(13) & 0.9364(12) & 0.9390(15) & 0.9436(13) \\
    0.5   & 1     & 0.8982(13) & 0.9136(14) & 0.9273(14) & 0.9361(15) & 0.9421(14) & 0.9432(13) & 0.9480(15) \\
    0.6   & 1.22  & 0.9000(13) & 0.9157(15) & 0.9300(13) & 0.9383(14) & 0.9470(13) & 0.9484(14) & 0.9538(14) \\
    0.7   & 1.53  & 0.9024(12) & 0.9195(15) & 0.9341(13) & 0.9438(15) & 0.9508(15) & 0.9532(15) & 0.9573(13) \\
    0.8   & 2     & 0.9037(14) & 0.9210(12) & 0.9375(13) & 0.9452(15) & 0.9530(13) & 0.9574(14) & 0.9616(15) \\
    0.8605 & 2.48  & 0.9042(14) & 0.9205(14) & 0.9381(16) & 0.9477(15) & 0.9560(13) & 0.9591(16) & 0.9641(11) \\
    0.9   & 3     & 0.9040(11) & 0.9238(12) & 0.9375(13) & 0.9480(15) & 0.9562(15) & 0.9598(12) & 0.9666(15) \\
\hline
    \end{tabular}%
  \label{tab:KS99}%
     \end{threeparttable}
   \end{sideways}
   \vfill
\endgroup
    \leavevmode\vfill
    \begingroup
\vfill
\hfil
\centering
\small
\begin{sideways}
  \begin{threeparttable}
  \centering
  \caption{left-truncated log-logistic AD($A^2$) significance level  85\%}
 \begin{tabular}{|rr|ccccccc|}
\hline
    p    & $\sqrt{\eta}$ & 30 & 50 & 100 & 200 & 500 & 1000 & 10000 \\
\hline
    0     & 0     & 0.4998(07) & 0.5010(07) & 0.5011(07) & 0.5013(07) & 0.5021(07) & 0.5021(07) & 0.5021(07) \\
    0.0323 & 0.18  & 0.4973(07) & 0.4974(07) & 0.4964(07) & 0.4968(08) & 0.4969(06) & 0.4976(07) & 0.4962(06) \\
    0.1   & 0.33  & 0.5031(07) & 0.5030(07) & 0.5040(06) & 0.5032(07) & 0.5035(07) & 0.5038(07) & 0.5037(07) \\
    0.2   & 0.5   & 0.5120(07) & 0.5125(07) & 0.5133(07) & 0.5131(06) & 0.5126(07) & 0.5142(07) & 0.5137(08) \\
    0.3   & 0.65  & 0.5190(07) & 0.5197(08) & 0.5209(08) & 0.5213(08) & 0.5218(07) & 0.5220(07) & 0.5216(07) \\
    0.4   & 0.82  & 0.5233(07) & 0.5266(07) & 0.5286(08) & 0.5289(07) & 0.5289(08) & 0.5292(08) & 0.5294(08) \\
    0.5   & 1     & 0.5283(07) & 0.5314(09) & 0.5340(07) & 0.5353(08) & 0.5356(07) & 0.5357(08) & 0.5357(08) \\
    0.6   & 1.22  & 0.5309(08) & 0.5352(07) & 0.5384(07) & 0.5406(08) & 0.5413(09) & 0.5406(08) & 0.5413(08) \\
    0.7   & 1.53  & 0.5332(07) & 0.5379(09) & 0.5419(08) & 0.5442(08) & 0.5457(09) & 0.5461(09) & 0.5460(08) \\
    0.8   & 2     & 0.5349(08) & 0.5401(08) & 0.5453(08) & 0.5462(08) & 0.5496(08) & 0.5516(08) & 0.5508(08) \\
    0.8605 & 2.48  & 0.5361(08) & 0.5404(09) & 0.5461(08) & 0.5483(08) & 0.5516(09) & 0.5536(08) & 0.5537(09) \\
    0.9   & 3     & 0.5360(09) & 0.5422(08) & 0.5462(09) & 0.5497(09) & 0.5526(09) & 0.5541(10) & 0.5549(09) \\
\hline
    \end{tabular}%
  \label{tab:AD85}%
  \centering
\caption{left-truncated log-logistic AD($A^2$) significance level 90\%}
 \begin{tabular}{|rr|ccccccc|}
\hline
    p    & $\sqrt{\eta}$ & 30 & 50 & 100 & 200 & 500 & 1000 & 10000 \\
\hline
    0     & 0     & 0.5589(08) & 0.5601(09) & 0.5608(08) & 0.5616(09) & 0.5620(08) & 0.5629(08) & 0.5628(09) \\
    0.0323 & 0.18  & 0.5568(08) & 0.5570(09) & 0.5560(08) & 0.5570(09) & 0.5572(08) & 0.5574(09) & 0.5561(09) \\
    0.1   & 0.33  & 0.5643(09) & 0.5641(09) & 0.5657(09) & 0.5647(09) & 0.5648(09) & 0.5657(09) & 0.5651(09) \\
    0.2   & 0.5   & 0.5756(08) & 0.5759(10) & 0.5768(09) & 0.5763(09) & 0.5757(09) & 0.5776(09) & 0.5769(10) \\
    0.3   & 0.65  & 0.5839(09) & 0.5848(10) & 0.5865(10) & 0.5868(10) & 0.5875(09) & 0.5873(10) & 0.5874(09) \\
    0.4   & 0.82  & 0.5897(09) & 0.5933(10) & 0.5955(10) & 0.5956(09) & 0.5957(09) & 0.5960(10) & 0.5971(11) \\
    0.5   & 1     & 0.5960(09) & 0.5987(10) & 0.6025(09) & 0.6043(10) & 0.6040(09) & 0.6043(11) & 0.6039(10) \\
    0.6   & 1.22  & 0.5985(09) & 0.6040(11) & 0.6076(11) & 0.6096(09) & 0.6112(11) & 0.6106(10) & 0.6104(10) \\
    0.7   & 1.53  & 0.6022(09) & 0.6070(10) & 0.6122(10) & 0.6155(09) & 0.6165(11) & 0.6174(11) & 0.6169(10) \\
    0.8   & 2     & 0.6038(11) & 0.6102(11) & 0.6165(11) & 0.6173(10) & 0.6220(12) & 0.6234(11) & 0.6224(10) \\
    0.8605 & 2.48  & 0.6056(09) & 0.6106(11) & 0.6171(10) & 0.6201(11) & 0.6235(11) & 0.6255(11) & 0.6261(11) \\
    0.9   & 3     & 0.6053(11) & 0.6123(10) & 0.6170(11) & 0.6218(11) & 0.6247(10) & 0.6268(12) & 0.6278(11) \\
\hline
    \end{tabular}%
  \label{tab:AD90}%
  \centering
  \caption{left-truncated log-logistic AD($A^2$) significance level 99\%}
    \begin{tabular}{|rr|ccccccc|}
\hline
    p    & $\sqrt{\eta}$ & 30 & 50 & 100 & 200 & 500 & 1000 & 10000 \\
\hline
    0     & 0     & 0.8917(29) & 0,8956(30) & 0,8981(28) & 0,9035(30) & 0,8998(31) & 0,9050(28) & 0,9027(32) \\
    0,0323 & 0,18  & 0,8962(27) & 0,8975(32) & 0,8959(29) & 0,8984(30) & 0,8979(29) & 0,8984(28) & 0,8999(29) \\
    0,1   & 0,33  & 0,9199(30) & 0,9163(32) & 0,9210(32) & 0,9177(33) & 0,9168(32) & 0,9183(31) & 0,9161(34) \\
    0,2   & 0,5   & 0,9473(32) & 0,9461(34) & 0,9462(31) & 0,9424(35) & 0,9407(30) & 0,9422(34) & 0,9465(33) \\
    0,3   & 0,65  & 0,9716(38) & 0,9701(37) & 0,9657(35) & 0,9656(30) & 0,9682(26) & 0,9688(33) & 0,9657(34) \\
    0,4   & 0,82  & 0,9831(33) & 0,9893(37) & 0,9893(35) & 0,9864(33) & 0,9885(30) & 0,9885(35) & 0,9853(32) \\
    0,5   & 1     & 0,9979(35) & 1,0001(39) & 1,0035(33) & 1,0043(36) & 1,0021(33) & 1,0008(38) & 1,0015(32) \\
    0,6   & 1,22  & 1,0017(38) & 1,0091(33) & 1,0162(34) & 1,0176(36) & 1,0211(36) & 1,0226(35) & 1,0201(36) \\
    0,7   & 1,53  & 1,0085(35) & 1,0165(39) & 1,0235(33) & 1,0301(37) & 1,0362(40) & 1,0340(29) & 1,0347(38) \\
    0,8   & 2     & 1,0172(37) & 1,0212(37) & 1,0317(35) & 1,0357(39) & 1,0437(34) & 1,0474(31) & 1,0438(37) \\
    0,8605 & 2,48  & 1,0154(37) & 1,0241(38) & 1,0343(40) & 1,0452(40) & 1,0485(39) & 1,0544(35) & 1,0524(37) \\
    0,9   & 3     & 1,0180(36) & 1,0319(34) & 1,0350(34) & 1,0436(40) & 1,0523(36) & 1,0540(32) & 1,0605(32) \\
\hline
    \end{tabular}%
  \label{tab:AD99}%
    \end{threeparttable}
   \end{sideways}
   \vfill
\endgroup

\newpage
\normalsize
\section{Proof of statement (2) in Lemma \ref{thm:bounds}.}\label{sec:ineq2}

We follow the proof of \cite{Gupta1999} and adapt their notation and define the quantities $a=-\ln{\lambda}$, $s_i=\ln{X_i}$ 
and 
$S=\sum_{i=1}^N \ln{X_i}=\sum_{i=1}^N s_i$.  We write now for simplicity $\varphi(a,\beta)$ rather than $\varphi(\lambda,\beta)$ and use the fact that all $s_i$ are positive because of the left-truncation equals 1. First, however, we start with two useful Propositions: 

\begin{myprop}{\label{thm:trivial}}
For any $m>0$ and $c \in  \mathbb{R}$ let $A\in \mathbb{R}$ satisfy the following inequalities
\begin{enumerate}[{(i)}]
\item $A \leq m c$
\item $A \leq - m c$
\end{enumerate}
Then we have the following inequality
\begin{eqnarray}
A \leq - m |c|
\nonumber
\end{eqnarray}
\end{myprop}
{\bf Proof:}

We have $A\leq 0$ by adding up inequalities (i) and (ii). Consider first the case when $c>0$ in which case inequality (ii) is tighter than inequality (i), thus $A\leq -m c = -m |c|$. Likewise when $c<0$ the inequality (i) is tighter than inequality (ii), thus $A\leq m c = m (-|c|)$. The case $c=0$ is trivial.  $\blacksquare$

\begin{myprop}{\label{thm:Jensen}}
For $z_i \in \mathbb{R}$ we have the following inequalities
\begin{eqnarray}
\frac{1}{N}\sum_{i=1}^N z_i  \leq  \ln{\left(1+ \exp{\left[\frac{1}{N}\sum_{i=1}^N z_i\right]} \right)}  \leq \frac{1}{N}\sum_{i=1}^N \ln{\left( 1 + e^{z_i} \right)}\quad.
\nonumber
\end{eqnarray}
\end{myprop}

{\bf Proof:}

Note that the function $\psi(z)=\ln \left(1+\exp{[z]}\right)$ is convex because $\psi''(z)= e^z/(1+e^z)^2>0$ for all $z\in \mathbb{R}$. The right-hand side inequality is Jensen's inequality (e.g. \cite{Hardy1952}, p.~74 Eq. (3.8.1) and (3.8.2)) whereas the left-hand side follows from monotonicity of the logarithm. $\blacksquare$

{\bf Proof of statement (2) in Lemma \ref{thm:bounds}.}

In our new notation the objective function Eq.~(\ref{eq:objective}) reads as 
\begin{equation}
\varphi(a,\beta)
=N \ln{ \left(1+e^a  \right)} + N \ln{\beta} + N a + \beta S
-2 \sum_{i=1}^{N}  \ln{ \left[ 1+ e^{a+\beta s_i}   \right] }\quad.
\tag{\ref{eq:objective}}
\end{equation}
We shall derive from this equation various inequalities. Because the last term is negative we have immediately the following inequality
\begin{eqnarray}
\varphi(a,\beta)
& \leq & N \ln{ \left(1+e^a  \right)} + N \ln{\beta} + N a + \beta S\quad.
\label{eq:ineq1}
\end{eqnarray}
Using the entire inequality chain from Proposition \ref{thm:Jensen}. we obtain another inequality from Eq.~(\ref{eq:objective})
\begin{eqnarray}
\varphi(a,\beta)
& \leq & N \ln{ \left(1+e^a  \right)} + N \ln{\beta} - N a - \beta S\quad.
\label{eq:ineq2}
\end{eqnarray}
Applying Proposition \ref{thm:trivial}. to the inequalities Eq.~(\ref{eq:ineq1}) and (\ref{eq:ineq2}) we get
\begin{eqnarray}
\varphi(a,\beta)
& \leq & N \ln{ \left(1+e^a  \right)} + N \ln{\beta} - \beta N \left| \frac{a}{\beta} + \frac{S}{N} \right| \quad.
\label{eq:ineq3}
\end{eqnarray}
A final upper bound is obtained from Eq.~(\ref{eq:objective}) using Proposition \ref{thm:Jensen}.  for any $i=1, 2, ..., N$
\begin{eqnarray}
\varphi(a,\beta)
&\leq & N \ln{ \left(1+e^a  \right)}  + N \ln{\beta}
- \sum_{i=1}^{N}  \ln{ \left[ 1+ e^{a+\beta s_i}   \right] }
\nonumber \\
& \leq & N \ln{ \left(1+e^a  \right)}  + N \ln{\beta} - \beta \left( \frac{a}{\beta} + s_i \right)\quad.
\label{eq:ineq4}
\end{eqnarray}

{\bf Case I:} Consider first $a\geq0$. We can bound the positive quantity $\ln{(1+e^a)}$ above by $\ln{2} + a$ and thus inequality Eq.~(\ref{eq:ineq2}) yields altogether
\begin{eqnarray}
\varphi(a,\beta)
& \leq & N \ln{2} + N \ln{\beta} - \beta S
\label{eq:bound1}
\end{eqnarray}
the desired upper bound independent of $a$.

 {\bf Case II:} Consider now $a<0$ We can bound the positive quantity $\ln{(1+e^a)}$ above by $\ln{2}$.
In this part we shall use the inequality Eq.~(\ref{eq:ineq3}) reading as
\begin{eqnarray}
\varphi(a,\beta)
& \leq & N \ln{2} + N \ln{\beta} - \beta N \left| \frac{a}{\beta} + \frac{S}{N} \right|
\label{eq:ineq3_v2}
\end{eqnarray}
and the inequality Eq.~(\ref{eq:ineq4}) reading as
\begin{eqnarray}
\varphi(a,\beta)
& \leq & N \ln{2}  + N \ln{\beta} - \beta \left( \frac{a}{\beta} + s_i \right)\quad.
\label{eq:ineq4_v2}
\end{eqnarray}

 Because not all $s_i$ are equal there exists an index $i_0$ with 
\begin{eqnarray}
s_{i_0} - \frac{S}{N} = \varepsilon_0 > 0\quad.
\nonumber
\end{eqnarray}
We have $(-\infty, s_{i_0} - \varepsilon_0) \cup (S/N+\varepsilon_0,+\infty) = \mathbb{R}$. Again we consider 2 sub-cases. 
If we have  $-a/\beta \in (S/N+\varepsilon_0,+\infty)$ then
\begin{eqnarray}
-\frac{a}{\beta} > \frac{S}{N}+ \varepsilon_0
\label{eq:subcaseA}
\end{eqnarray}
from which we obtain
\begin{eqnarray}
\left| \frac{a}{\beta} + \frac{S}{N} \right| =- \left( \frac{a}{\beta} + \frac{S}{N}\right) > \varepsilon_0\quad.
\label{eq:subcaseA_v2}
\end{eqnarray}
Now we use this lower bound Eq.~(\ref{eq:subcaseA_v2}) in conjunction with inequality Eq.~(\ref{eq:ineq3_v2}) to obtain
\begin{eqnarray}
\varphi(a,\beta)
& \leq & N \ln{ 2} + N \ln{\beta} - \beta N \varepsilon_0\quad.
\label{eq:bound2A}
\end{eqnarray}
Likewise if we have $-a/\beta \in (-\infty,s_{i_0}-\varepsilon_0)$ then
\begin{eqnarray}
-\frac{a}{\beta} < s_{i_0}-\varepsilon_0
\label{eq:subcaseB}
\end{eqnarray}
from which we obtain
\begin{eqnarray}
\varepsilon_0 < s_{i_0}+ \frac{a}{\beta} \quad.
\label{eq:subcaseB_v2}
\end{eqnarray}
Now we use this lower bound Eq.~(\ref{eq:subcaseB_v2}) in conjunction with inequality Eq.~(\ref{eq:ineq4_v2})
\begin{eqnarray}
\varphi(a,\beta)
& \leq & N \ln{2}  + N \ln{\beta} - \beta \varepsilon_0\quad.
\label{eq:bound2B}
\end{eqnarray}
Hence for any $a<0$ we obtain from Eq.~(\ref{eq:bound2A}) and (\ref{eq:bound2B}) the desired bound independent of $a$
\begin{eqnarray}
\varphi(a,\beta)
& \leq & N \ln{2}  + N \ln{\beta} - \beta \varepsilon_0
\label{eq:bound2}
\end{eqnarray}
since $N\geq 2$.

From Case I Eq.~(\ref{eq:bound1}) and Case II Eq.~(\ref{eq:bound2}) we obtain altogether
\begin{eqnarray}
\varphi(a,\beta)
& \leq & N \ln{2}  + N \ln{\beta} - \beta \min{\{S, \varepsilon_0\}}
\label{eq:bound2}
\end{eqnarray}
and thus the existence of a sufficiently large $\beta_2>\beta_1>0$, independent of $a$ such that $\varphi(a,\beta)<-M$ for $\beta>\beta_2$ and all $a\in \mathbb{R}$. This concludes the proof. $\blacksquare$


\begin{thebibliography}{widest entry}

\bibitem{Fisk1961}
Fisk, P.R. (1961) "The Graduation of Income Distributions", \emph{Econometrica} \textbf{29(2)}, pp. 171-185. 

\bibitem{Shoukri1988}
Shoukri, M.M., Mian, I.U.H. and  Tracy, D.S. (1988) "Sampling properties of estimators of the log-logistic distribution with application to Canadian precipitation data", \emph{The Canadian Journal of Statistics}  \textbf{16 (3)}, pp. 223-236.

\bibitem{Sinclair1988}
Ahmad, M.I., Sinclair, C.D. and Werritty, A. (1988) "Log-logistic flood frequency analysis", \emph{Journal of Hydrology} \textbf{98}, pp. 205-224

\bibitem{Reath2018}
Reath, J. , Dong, J. and Wang, M. (2018) "Improved parameter estimation of the log-logistic distribution with applications", \emph{Computational Statistics}, \textbf{33 (1),} pp 339-356

\bibitem{He2020}
He, X. , Chen, W. , Qian,W. (2020) "Maximum likelihood estimators of the parameters of the log-logistic distribution", \emph{Statistical Papers} \textbf{61(5)}, pp 1875-1892

\bibitem{Kendall1979} 
Kendall, M. and Stuart, A.  (1979) "The advanced theory of statistics II - Inference and relationship", 4th revised edition, Griffin London.

\bibitem{Cohen1991}
Cohen, C. (1991) "Truncated and Censored Samples -Theory and Applications", 1st edition, CRC Press Boca Raton

\bibitem{Guscott2018}
Guscott, J.C. (2018) "Reliable Statistical Methods and their Applications for Testing Incomplete Multidisciplinary Data", MPhil Thesis, Physics Department, University of Adelaide, South Australia,
{\url{https://hdl.handle.net/2440/127169}}

\bibitem{Castillo1994}
Castillo, J. del (1994) "The singly truncated normal distribution: A non-steep exponential family", \emph{Annals of the Institute of Statistical Mathematics} \textbf{45 (1)}, pp. 57-66

\bibitem{KreerKizilersuThomasReis2015}
Kreer, M., Kizilersu A., Thomas A.W. and dos Reis A.E. (2015). 
"Goodness-of-fit tests and applications for left-truncated
  Weibull distributions to non-life insurance."  \emph{European Actuarial Journal}, \textbf{5 (1)}, pp 139-163.


\bibitem{Gupta1999}
Gupta, R.C., Akman, O. and Levin, S. (1999) "A study of log-logistic model in survival analysis", \emph{Biometrical Journal} \textbf{41}, pp 431-443.

\bibitem{Antle1970}
Antle, C.E., Klimko, L. and Harkness, W.  (1970) "Confidence intervals for the parameters of the logistic distribution", \emph{Biometrika} \textbf{57 (2)}, 397-402.

\bibitem{Kolmogorov1933}
Kolmogorov, A. (1933) "Sulla detenninazione empirica di una legge di distribuzione" , \emph{Giornale
dell’Istituto Italiano degli Attuari}, \textbf{4}, pp. 83–91.

\bibitem{Smirnov1948}
Smirnov, N. (1948) "Table for estimating the goodness of fit of emperical distributions", \emph{The Annals of Mathematical Statistics},  \textbf{19(2)}, pp. 279–281.

\bibitem{AndersonDarling1952}
Anderson, T.W. and Darling, D.A. (1952) "Asymptotic theory of certain ’goodness of
fit’ criteria based on stochastic processes", \emph{The Annals of Mathematical Statistics},
\textbf{23(2)}, pp. 193–212.

\bibitem{Mosteller1946} 
Mosteller, F. (1946). "On some useful inefficient statistics". \emph{The Annals of Mathematical Statistics} \textbf{ 17  (4)}, pp 377-–408.

\bibitem{Cuddington2011}
Cuddington, J.T. and Navidi, W. (2011) "A critical assessment of simulated critical values"
\emph{Communications in Statistics-Simulation and Computation}, \textbf{40}, pp. 719–727.

\bibitem{Stephens1979}
Stephens, M.A. (1979) "Test of fit for the logistic distribution based on the empirical distribution function", \emph{Biometrika} \textbf{66 (3)} pp. 591-595

\bibitem{Lee2003}
Lee, E.T. and Wang, J. W.  (2003) "Statistical Methods for Survival Data Analysis", 3rd edition, John Wiley \& Sonds Inc. Hoboken, New Jersey.  

\bibitem{AlShomrani2016}
Al-Shomrani, A. A, Shawky, A.I.,  Arif, O.H. and Aslam, M. (2016) "Log-logistic distribution for survival data analysis using MCMC", \emph{SpringerPlus} \textbf{5}(1), pp 1774–1790 .

\bibitem{Ashkar2006}
Ashkar, F. and Mahdi, S. (2006) "Fitting the log-logistic distribution by generalized moments", \emph{ Journal of Hydrology}, {\bf{328}}, pp 694--703.

\bibitem{DWD2022}
Climate data base of the DWD Deutscher Wetterdienst (German Meteorology Office), Offenbach Germany,
{\url{http://opendata.dwd.de/climate_environment/CDC/regional_averages_DE/annual/precipitation/regional_averages_rr_year.txt}}

\bibitem{Toronto2022}
Toronto Weather Statistics (2022), "Total Precipitation - Annual data Jul 1-Jun 30 for Toronto",
 {\url{toronto.weatherstats.ca/charts/precipitation-wyearly.html}}

\bibitem{Adelaide2022}
{\url{http://www.bom.gov.au/jsp/ncc/cdio/weatherData/av?p_nccObsCode=136&p_display_type=dailyDataFile&p_startYear=1884&p_c=-106001191&p_stn_num=023011}}
{\url{http://www.bom.gov.au/climate/data/index.shtml}}

\bibitem{Wingo1983}
Wingo, D. R.  (1983) ``Maximum likelihood methods for fitting the Burr type XII distribution to life test data", \emph{Biometrical Journal} \textbf{Vol. 25} (1), pp. 77-84. 

\bibitem{Shao2004}
Shao, Q. (2004) ``Notes on maximum likelihood estimation for the three-parameter Burr XII distribution", \emph{Computational Statistics $\&$ Data Analysis}, \textbf{45} (3) pp 675 – 687.

\bibitem{Wang2010}
Wang ,  F.-K. and Cheng Y.-F. (2010) ``Robust regression for estimating the Burr XII parameters with outliers", \emph{Journal of Applied Statistics} \textbf{37} (5), pp. 807-819.

\bibitem{Ostrowski1960}
Ostrowski, A.M. (1960), "Solution of Equations and Systems of Equations", Academic Press New York and London.

\bibitem{Hardy1952}
Hardy, G., Littlewood, J.E. and Polya, G.  "Inequalities", 2nd edition reprinted 1994, Cambridge University Press, Cambridge.

\bibitem{Martin1987}
Harris, G. and Martin, C.  (1987). "Shorter notes: The roots of a polynomial vary continuously as a function of the coefficients." \emph{Proceedings of the American Mathematical Society} \textbf{100 (2)}, pp 390--392.

\bibitem{AntleBain1969}
Antle, C.E. and Bain, L.J.  (1969) "A property of maximum likelihood estimators of location and scale parameters", \emph{SIAM Review} \textbf{11 (2)}, pp. 251-253 .

\end{thebibliography}
\end{document}